\documentclass[twocolumn,aps,superscriptaddress,showpacs,floatfix,prd,noshowpacs]{revtex4-2}
\usepackage[body={19.cm,24.cm}]{geometry}     
\usepackage{mathrsfs}
\usepackage{amssymb}
\usepackage{amsmath}
\usepackage{graphicx}
\usepackage[normalem]{ulem}
\usepackage{xcolor}
\usepackage{bm}
\usepackage{longtable}
\usepackage{slashed}
\usepackage{enumitem}
\usepackage{empheq}
\usepackage{booktabs}
\usepackage{multirow}

\usepackage{titlesec}
\renewcommand{\thesection}{\Roman{section}}
\titleformat{\section}{\small\bfseries\centering}{\thesection.}{0.5em}{}

\usepackage[T1]{fontenc}
\usepackage[utopia]{mathdesign}

\usepackage[breaklinks=true]{hyperref}
\hypersetup{
  colorlinks=true,
  citecolor=magenta,
  linkcolor=black,
  urlcolor=teal,
}

\renewcommand\sout{\bgroup \color{red} \ULdepth=-.5ex \ULset}

\renewcommand{\v}[1]{\textbf{#1}}
\renewcommand{\rm}[1]{\textrm{#1}}
\renewcommand{\d}{\mathrm{d}}

\usepackage{tikz}

\definecolor{lime}{HTML}{A6CE39}
\DeclareRobustCommand{\orcidicon}{
	\begin{tikzpicture}
	\draw[lime, fill=lime] (0,0) 
	circle [radius=0.16] 
	node[white] {{\fontfamily{qag}\selectfont \tiny ID}};
	\draw[white, fill=white] (-0.0625,0.095) 
	circle [radius=0.007];
	\end{tikzpicture}
	\hspace{-2mm}
}
\foreach \x in {A, ..., Z}{%
	\expandafter\xdef\csname orcid\x\endcsname{\noexpand\href{https://orcid.org/\csname orcidauthor\x\endcsname}{\noexpand\orcidicon}}
}

\DeclareSymbolFont{cmsymbols}{OMS}{cmsy}{m}{n}
\DeclareSymbolFontAlphabet{\mathcal}{cmsymbols}

\begin{document}

%\title{A New Scaling Probe of the Neutron Star Core Equation of State from GW170817}
\title{A New Scaling of Neutron Star Tidal Deformability for Directly Probing\\ the Central Equation of State Parameter}

\author{Jian-Hao Shi\orcidD{}}
\affiliation{Department of Physics and Center for Field Theory and Particle Physics, Fudan University, Shanghai 200438, China} 
\affiliation{Shanghai Research Center for Theoretical Nuclear Physics, NSFC and Fudan University, Shanghai 200438, China}
\author{Bao-Jun Cai\orcidA{}}\email{bjcai@fudan.edu.cn}
\affiliation{Shanghai Research Center for Theoretical Nuclear Physics, NSFC and Fudan University, Shanghai 200438, China}
\affiliation{Key Laboratory of Nuclear Physics and Ion-beam Application (MOE), Institute of Modern Physics, Fudan University, Shanghai 200433, China} 
\author{Bao-An Li\orcidB{}}\email{Bao-An.Li$@$etamu.edu}
\affiliation{Department of Physics and Astronomy, East Texas A\&M University, Commerce, TX 75429-3011, USA}
\author{Yu-Gang Ma\orcidC{}}\email{mayugang$@$fudan.edu.cn}
\affiliation{Shanghai Research Center for Theoretical Nuclear Physics, NSFC and Fudan University, Shanghai 200438, China}
\affiliation{Key Laboratory of Nuclear Physics and Ion-beam Application (MOE), Institute of Modern Physics, Fudan University, Shanghai 200433, China} 
\affiliation{College of Physics, East China Normal University, Shanghai 200241, China}

\date{\today}

\newcommand{\x}{\mathrm{X}}
\newcommand{\y}{\mathrm{Y}}
\newcommand{\hr}{\widehat{r}}
\newcommand{\hP}{\widehat{P}}
\newcommand{\heps}{\widehat{\varepsilon}}
\newcommand{\hrho}{\widehat{\rho}}

\begin{abstract}
The dimensionless tidal deformability, $\Lambda$, of neutron stars (NSs), inferred from gravitational-wave (GW) observations, has thus far been used primarily to constrain the pressure of dense matter near twice nuclear saturation density, leaving the core equation of state (EOS) largely inaccessible to inspiral-phase GW observations. We show that a fundamental central EOS parameter can be probed directly through $\Lambda$ using a perturbative analysis of the dimensionless stellar-structure and tidal-response equations formulated in terms of scaled intrinsic variables, without invoking any specific EOS model. We uncover a remarkable EOS-insensitive scaling relation between $\Lambda$ and the central EOS parameter $\x\equiv P_{\rm c}/\varepsilon_{\rm c}$, where $P_{\rm c}$ and $\varepsilon_{\rm c}$ denote the central pressure and energy density, respectively. The relation is validated against a broad ensemble of physically viable EOSs. Applying it to tidal deformabilities inferred from events such as GW170817 enables a direct determination of $\x$. 
By adopting the previously established effective upper bound $\x\lesssim0.381$, we infer a tight lower bound, $\Lambda_{\rm{TOV}}\gtrsim9.1^{+0.5}_{-0.5}$, for maximum-mass NSs along stable mass-radius sequences, quantitatively demonstrating that even the most compact stable NSs remain distinctly separated from black holes, for which $\Lambda_{\rm{BH}}=0$.
These findings reveal a previously unrecognized connection between inspiral-phase tidal deformability and the central EOS parameter $\x$, establishing a direct link between GW observables and the microphysics of ultradense matter in the strong-gravity regime. The resulting scaling establishes inspiral-phase tidal deformability as a direct and largely model-insensitive probe of the central EOS parameter $\x$, and hence of key properties of NS cores.
\end{abstract}

\maketitle

\fontdimen2\font=2.pt

\section{Introduction}

The discovery of GW170817 demonstrated that the dimensionless tidal deformability $\Lambda$, encoded in the inspiral gravitational-wave (GW) signal from neutron-star (NS) mergers, carries invaluable information about NS interiors\,\cite{Abbott2017,Abbott2018}. Together with recent measurements of NS masses $M_{\rm{NS}}$ and radii $R$\,\cite{Riley19,Miller19,Fon21,Riley21,Miller21,Salmi22,Choud24,Reardon24,Mauv25}, tidal-deformability observations have significantly advanced our understanding of the equation of state (EOS) of dense matter; see, e.g., Refs.\,\cite{Baym18,Isa18,LCXZ21,Latt21,Oertel17,Bai19,Dri21,Lovato22,Soren23,Chat24,Alar25,LiA25,Wat16} for recent reviews. In particular, owing to its strong sensitivity to stellar radius\,\cite{LP01}, $\Lambda$ has emerged as a powerful probe of matter at densities around twice nuclear saturation density\,\cite{Farr13,De18,Lim18,Zhao18,Most18,Zhou19,Li2019,Chat20,Ferr24,Ferr25,ZhouX24,Tsang24,Mroczek24,Ripley24,Malik24,CuiY25}. However, little model-independent information about the EOS at substantially higher densities has been extracted from currently available tidal-deformability measurements using traditional analyses based on the stellar-structure and tidal-response equations. Meanwhile, several studies have suggested that future observations of post-merger high-frequency GWs may probe the high-density EOS more directly, see, e.g., Refs.\,\cite{Hot2011,Bau2012,Rez2016,Bau2016,Mos2019}. While awaiting such measurements from next-generation GW detectors, it is natural to ask whether inspiral-phase tidal-deformability data already contain previously unrecognized information about the EOS deep inside NS cores.

Current efforts to infer the high-density EOS generally rely on parameterized EOS models. In these approaches, information obtained at low and intermediate densities from nuclear many-body theories, terrestrial experiments, and astrophysical observations is extrapolated to higher densities, and the resulting constraints inevitably inherit some degree of model dependence. Although tidal deformability has become one of the most important observables for constraining dense matter properties, its implications for the core EOS remain indirect and are often entangled with assumptions about the underlying EOS parameterization. Whether tidal deformability itself encodes robust information about the EOS deep inside NSs therefore remains an open question.

A key step toward answering this question is to determine whether the relativistic tidal-response equation possesses intrinsic scaling properties. In this work, we demonstrate that such a scaling indeed exists. Building upon the recently developed IPAD-TOV (Intrinsic and Perturbative Analyses of the Dimensionless TOV equations) framework \cite{CaiLi2025Review}, which revealed intrinsic scaling relations \cite{CaiLiZhang2023ApJ,CaiLiZhang2023PRD,CaiLi2024SSS,CaiLi2025Trace,Cai2024Front,CaiLiMa2026phi,CaiLiMa2026X} emerging directly from the dimensionless Tolman--Oppenheimer--Volkoff (TOV) equations\,\cite{Tolman1939,Oppenheimer1939,Web99}, we extend the analysis to the relativistic tidal-response equation\,\cite{Hinderer2008} and derive a new scaling relation for the dimensionless tidal deformability $\Lambda$. We find that the corresponding scaling factor $D(\x\,,\Psi)$ is governed primarily by two quantities: the central EOS parameter 
$\x\ \equiv \phi_{\rm c}\equiv P_{\rm c}/\varepsilon_{\rm c}$ and the logarithmic stability slope $\Psi=2\mathrm{d}\ln M_{\rm{NS}}/\mathrm{d}\ln\varepsilon_{\rm c}$\,\cite{CaiLiMa2026phi}, where $P_{\rm c}$ and $\varepsilon_{\rm c}$ denote the central pressure and energy density, respectively. The parameter $\x\ $ is equivalent to the central trace-anomaly measure $\Delta_{\rm c}\equiv 1/3-\x\ $\,\cite{Fuji22}. It can also be interpreted as the average EOS stiffness accumulated from the stellar surface to the center\,\cite{Saes24,Marc24,Li26}. This nonlocal character naturally links global observables, such as the radius and tidal deformability, to the thermodynamic properties of matter deep in the stellar core.

To evaluate the effectiveness of the scaling relation, we construct broad ensembles of calibrated EOSs based on both the nonlinear Walecka model\,\cite{SW1986} and meta-model descriptions\,\cite{ZhangLi2018,ZhangLi2019,XieLi2020,ZhangLi2021}, constrained by current nuclear-theory calculations, heavy-ion collision data, and astrophysical observations. The resulting scaling is further tested against 33 representative microscopic EOSs from the literature, including models featuring phase transitions, smooth crossovers, hyperons, and deconfined quarks; see the CompOSE database\,\cite{Typel15} and Ref.\,\cite{Ofengeim24} for further details. We demonstrate that observed tidal deformabilities can be mapped directly onto the central EOS parameter $\x$, providing, to our knowledge, the first direct extraction of core-EOS information from tidal-deformability measurements. When combined with the previously established mass scaling, the new relation also yields constraints on the central pressure and energy density of canonical NSs and predicts a robust lower bound, $\Lambda_{\rm{TOV}}\gtrsim 9.1$, for maximum-mass stable NS configurations (i.e., TOV NSs). The implications of the scaling for the sound-speed profile are likewise explored. These findings reveal a previously unrecognized connection between the inspiral phase tidal deformability and the central EOS parameter, establishing inspiral-phase GW observations as a direct and largely model-insensitive probe of ultradense matter in NS cores.

The rest of this paper is organized as follows. In Section \ref{sec:ipadtov}, we present the theoretical framework and derive the scaling factor $D(\x,\Psi)$ associated with the dimensionless tidal deformability. Section \ref{sec:AI-1} investigates the $D$-$\Lambda$ scaling for canonical NSs, from which the central EOS parameter $\x$ can be directly inferred from the observed $\Lambda$. Section \ref{sec:AI-2} extends this analysis to TOV NSs, where a lower bound on $\Lambda_{\rm{TOV}}$ is obtained. In Section \ref{AI-3}, we further constrain the central EOS for canonical NSs by combining the mass and tidal-deformability scalings. A summary is given in Section \ref{sec:summary}, while technical details are presented in the appendices.

\section{Scaling for dimensionless tidal deformability from the IPAD-TOV approach}
\label{sec:ipadtov}

In this section, we derive the scaling relation for the dimensionless tidal deformability,
$
\Lambda={2k_2}/{3\xi^5},
$
directly from the underlying stellar-structure and tidal-response equations, thereby elucidating how EOS-related quantities govern $\Lambda$. Here $k_2$ is the quadrupolar tidal Love number, and $\xi\equiv M_{\rm{NS}}/R$ denotes the stellar compactness\,\cite{Hinderer2008}.

Since $\Lambda$ is dimensionless, its dependence on the EOS can only arise through dimensionless combinations of thermodynamic quantities. The most natural such quantity is the EOS parameter
$
\phi \equiv {P}/{\varepsilon},
$
which measures the local stiffness of matter. Unlike $\Lambda$ and the compactness $\xi$, both of which characterize the global structure of a NS, $\phi$ is a local quantity that varies throughout the stellar interior. One therefore expects $\Lambda$ to depend not on the local value of $\phi$ at a particular radius, but on an appropriately averaged EOS parameter that captures the integrated properties of the star. As we show below, the central value,
$
\x \equiv \phi_{\rm c}={P_{\rm c}}/{\varepsilon_{\rm c}}
$
which represents the energy-density-averaged sound speed squared (SSS) $s^2(\varepsilon)\equiv {\d P}/{\d\varepsilon}$ up to $\varepsilon_{\rm c}$ since
$\langle s^2(\varepsilon_\rm c) \rangle = \varepsilon_{\rm c}^{-1}\int_0^{\varepsilon_\rm c} s^2(\varepsilon')\d\varepsilon' = {P(\varepsilon_\rm c)}/{\varepsilon_\rm c}=\x$\,\cite{Saes24,Marc24,Li26}, provides such a measure and emerges naturally from the scaling analysis. Consequently, the tidal deformability may be expressed schematically as
$
\Lambda \sim D(\x,\ldots),
$
where $D$ is a dimensionless scaling function and the ellipsis denotes additional dimensionless quantities that may influence the stellar tidal response.
We emphasize here that $\x$ is not merely a local central quantity; through the TOV equations, it effectively encodes the integrated stiffness of the EOS from the surface to the center. Therefore, a global observable such as $\Lambda$ can correlate tightly with the core EOS parameter $\x$.

We first analyze the corresponding scaling behavior of the tidal Love number $k_2$, which is given by\,\cite{Hinderer2008}
\begin{align}
k_2 = &\frac{8}{5}\xi^5(1-2\xi)^2\left[2-y(\widehat{R})+2\xi\left(y(\widehat{R})-1\right)\right] \notag\\
&\times \left\{6\xi\left[2-y(\widehat{R})+\xi\left(5y(\widehat{R})-8\right)\right]\right.\notag \\
&+ 4\xi^3\left[13-11y(\widehat{R})+\xi\left(3y(\widehat{R})-2\right)+2\xi^2\left(1+y(\widehat{R})\right)\right] \notag\\
&\left.+ 3(1-2\xi)^2\ln(1-2\xi)\left[2-y(\widehat{R})+2\xi\left(y(\widehat{R})-1\right)\right]\right\}^{-1}.\label{def-k2}
\end{align}
The $k_2$ depends crucially on the dimensionless function $y(\hr)$ evaluated at the stellar surface, $y(\widehat{R})$, where $y(\hr)$ satisfies the differential equation\,\cite{Hinderer2008}
\begin{align}
&\hr \frac{\d y}{\d\hr} + y^2 + ye^{\lambda}\left[1+\hr^2\left(\hP-\heps\right)\right] \notag\\
&+ e^{\lambda}\left[\hr^2\left(5\heps + 9\hP + \frac{\heps+\hP}{\d\hP/\d\heps}\right) - 6\right] - \left(\hr\frac{\d\nu}{\d\hr}\right)^2=0.\label{eq:14}
\end{align}
Here, $\heps=\varepsilon/\varepsilon_{\rm{c}}$ and $\hP=P/\varepsilon_{\rm{c}}$ are the reduced energy density and pressure with respect to the central energy density $\varepsilon_{\rm{c}}$ in NSs. The reduced mass is defined as $\widehat{M}=M/Q$ within the dimensionless radius $\hr=r/Q$, where $Q=({4\pi \varepsilon_{\rm{c}}})^{-1/2}$\,\cite{CaiLi2025Review} sets the characteristic energy density/length scale (in units of $c=G=1$). In addition, $e^{\nu}$ and $e^{\lambda} = (1 - 2 \widehat{M}/\hr)^{-1}$ are respectively the (negative) time component and the radial component of the spacetime metric. In particular, the dimensionless metric function $\nu(\hr)$ is determined by the TOV equations\,\cite{Tolman1939,Oppenheimer1939}:
\begin{equation}
\frac{\d\hP}{\d\hr}=-\frac{\heps+{\hP}}{2}\frac{\d\nu}{\d\hr}
=- \frac{\heps\widehat{M} (1 + \hP/\heps) (1 + \hr^3 \hP/\widehat{M})}{\hr^2 (1 - 2\widehat{M}/\hr)};\frac{\d\widehat{M}}{\d\hr} = \hr^2 \heps.\label{eq:TOV1}
\end{equation}
To proceed analytically, we consider the near-center behavior of $y(\hr)$. It satisfies the boundary condition $y(0) = 2$, and we introduce the expansion $y(\hr) \approx 2 + \sum_{j=1}^{\infty}f_j\hr^j$. Substituting this expansion into Eq.\,\eqref{eq:14} shows that all odd-term coefficients vanish, and the leading nonzero contribution arises from $f_2$:
\begin{equation}
f_2 = -\frac{1 - 18 a_2 + 36 \x + 99 \x^2}{21 + 63\x},
\end{equation}
where $a_2$ is the expansion coefficient of the reduced energy density as $\heps\approx1+a_2\hr^2+\cdots$\,\cite{CaiLi2025Review}. Consequently, to this order one obtains $y(\widehat{R}) \approx 2 + f_2 \widehat{R}^2$.
The reduced radius squared $\widehat{R}^2$ is determined by the condition $\hP(\widehat{R})\approx0$, with $\hP\approx\x+b_2\hr^2+\cdots$, where $b_2$ is the leading non-trivial expansion coefficient\,\cite{CaiLiZhang2023ApJ}. The $\d\hP/\d\heps=\d P/\d\varepsilon$ in Eq.\,\eqref{eq:14} is the SSS, whose value at the NS center relates the coefficients $a_2$ and $b_2$ through $s_{\mathrm{c}}^2 = \d \hP_\rm c / \d \heps_\rm c = (\d \hP_\rm c / \d \hr) (\d \heps_\rm c / \d \hr)^{-1} = b_2/ a_2$\,\cite{CaiLi2025Review}.

The central SSS is determined from the derivative $\d M_{\rm{NS}}/\d\varepsilon_{\rm{c}}$ of the NS mass with respect to the central energy density:
\begin{align}
s_\mathrm{c}^2 \approx &\x \left( 1 + \frac{1 + \Psi}{3} \frac{1 + 3\x^2 + 4\x}{1 - 3\x^2}\right)\notag\\
&\times\left[1-\frac{24(1+\Psi)^2}{4+\Psi}\frac{\x}{50+37\Psi+5\Psi^2}\right],
\label{eq:sc2}
\end{align}
where $\Psi\equiv 2\d\ln M_{\rm{NS}}/\d\ln\varepsilon_{\rm{c}}$ denotes the log-stability slope\,\cite{CaiLiMa2026phi} of the NS mass with respect to the central energy density, $M_{\rm{NS}}\equiv \widehat{M}_{\rm{NS}}/Q\sim\widehat{M}_{\rm{NS}}/\sqrt{\varepsilon_{\rm{c}}}$ with $\widehat{M}_{\rm{NS}}\equiv\int_0^{\widehat{R}}\d\hr\hr^2\heps(\hr)$. In deriving Eq.\,\eqref{eq:sc2}, the contribution from $a_2\hr^2$ is included. For TOV NSs where $\Psi=0$, the correction term in Eq.\,\eqref{eq:sc2} reduces to $(1-3\x/25)$\,\cite{CaiLi2025Trace}.
For generally stable NSs below the TOV configuration on the mass-radius sequences, $\Psi>0$.
Likewise, the NS compactness $\xi$ can be shown to scale as
\begin{equation}
\xi={\widehat{M}_{\rm{NS}}}/{\widehat{R}}
\approx \frac{2\x}{1 + 3\x^2 + 4\x}\left[1+\frac{36(1+\Psi)}{50+37\Psi+5\Psi^2}\x\right]\equiv 2\Pi_{\rm c},
\label{eq:xi}
\end{equation}
here $2\x/(1+3\x^2+4\x)$ is the leading-order expression for the NS compactness\,\cite{CaiLiZhang2023ApJ}.
The correction term in Eq.\,\eqref{eq:xi} reduces to $1+18\x/25$\,\cite{CaiLi2025Trace} for $\Psi=0$.
It should be pointed out that the overall corrections in the square brackets of Eq.\,\eqref{eq:sc2} and Eq.\,\eqref{eq:xi} are of order $\x$, while higher-order contributions starting from $\x^2$ are expected to be small. 

Both the compactness $\xi$ and the central SSS $s_{\rm c}^2$ enter explicitly into the determination of $k_2$, as can be seen from its defining relation in Eq.\,\eqref{def-k2} as well as the corresponding contribution in Eq.\,\eqref{eq:14}. Consequently, the dependence of $k_2$ on $\x$ is intrinsically nontrivial.
Based on the expressions derived above, we obtain the generalized approximate expression for $k_2$ with the central EOS parameter $\x$ and $\Psi$ as the \begin{equation}
k_2(\x,\Psi)
\approx \frac{256}{5}
\frac{
\Pi_{\rm c}^5(1-4\Pi_{\rm c})^2
}{
H_2/H_1
+
3(1-4\Pi_{\rm c})^2\ln(1-4\Pi_{\rm c})
}.
\label{eq:k2}
\end{equation}
where the expressions for $H_1$ and $H_2$ are given in Appendix \ref{app:A}. 
Without surprise, $k_2$ depends on $\x$ in a very non-trivial manner 
because Eq.\,\eqref{eq:14} and the TOV equations are all highly nonlinear.
Taking $\Psi=0$ gives the corresponding $k_2$ for TOV NSs.

For small $\x$, one finds $k_2\approx \mathrm{const.}$, demonstrating that the quadrupolar tidal Love number saturates in the weak-gravity limit, where the tidal response is governed primarily by the shape of the density profile rather than by the overall stellar scale. Consequently, $k_2$ depends only weakly on the compactness, whereas the dimensionless tidal deformability $\Lambda$ is driven by the geometric factor $\xi^{-5}$, producing the characteristic Newtonian scaling $\Lambda\sim \xi^{-5}\sim\x^{-5}$.

We denote the quantity $2k_2/3\xi^5$ derived above by $D(\x,\Psi)$, where $k_2$ is given explicitly by Eq.~\eqref{eq:k2} and $\xi$ by Eq.~\eqref{eq:xi}. We refer to $D(\x,\Psi)$ as the \emph{intrinsic tidal-response function}, as it encapsulates the essential EOS dependence of the dimensionless tidal deformability $\Lambda$. In the following sections, we investigate the scaling relation between $D(\x,\Psi)$ and the tidal deformability $\Lambda$ obtained from exact numerical solutions of Eqs.~(\ref{eq:14}) and (\ref{eq:TOV1}). We demonstrate that $D$ serves as an intrinsic measure of the relativistic tidal response of NSs and provides the foundation for a direct mapping between tidal deformability and the EOS properties of NS cores.

\section{Extracting the central EOS-parameter $\x$ for canonical NSs straightforwardly via the $\Lambda$-$D$ scaling}
\label{sec:AI-1}

\begin{figure}[h!]
\centering
\includegraphics[width=7.7cm]{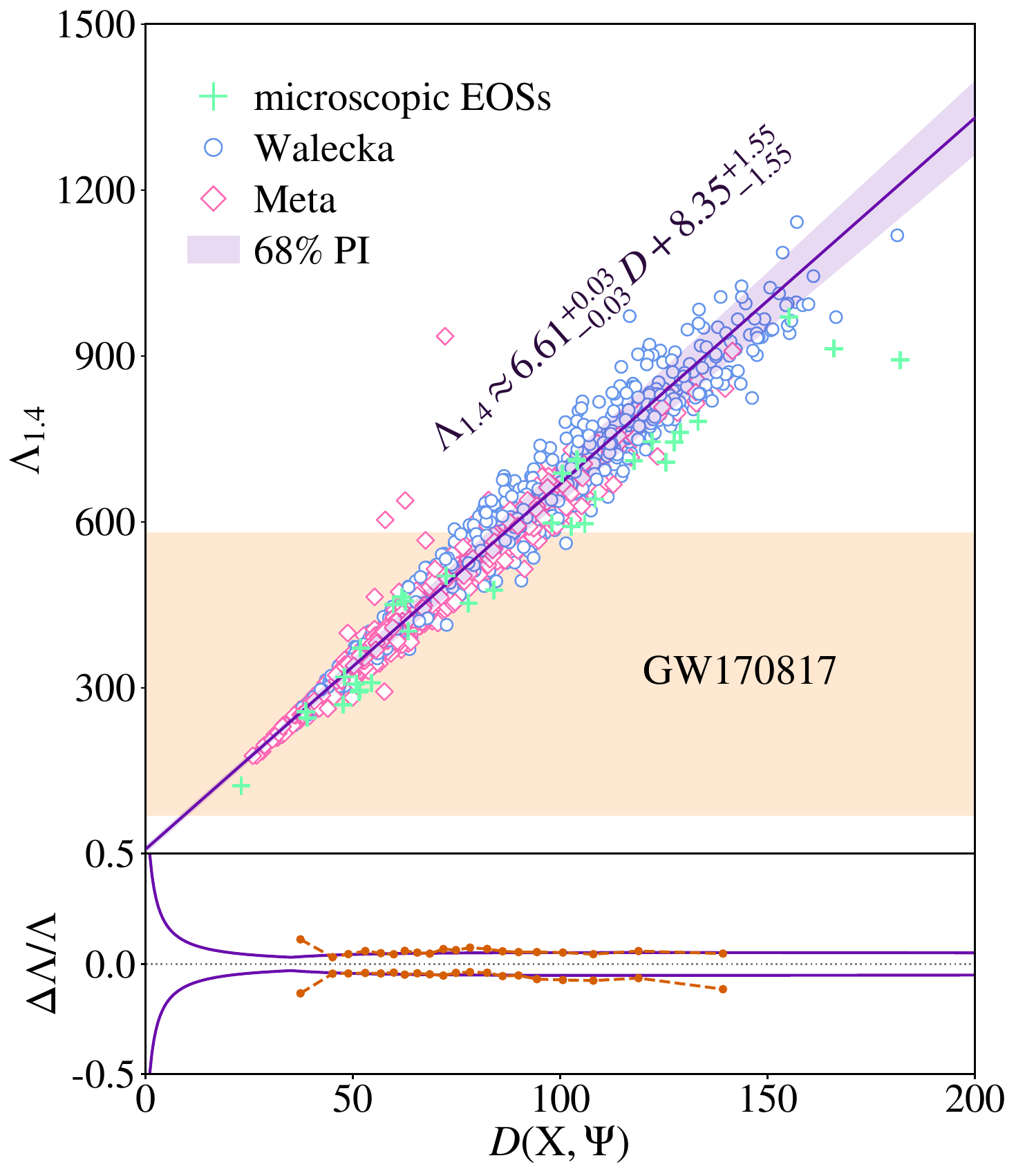}
\caption{(Color Online). Correlation between $\Lambda_{1.4}$ and the intrinsic tidal-response function $D(\x,\Psi)$ using the $2\times10^3$ EOS samples. The yellow band denotes the tidal deformability constraint inferred from GW170817, namely $70\lesssim \Lambda_{1.4}\lesssim580$. Points from extra 33 microscopic EOSs are shown by the green ``+'' symbol.
{The lower panel compares the FWLS 68\% prediction interval (purple) with the empirical 16th-84th percentile interval (orange) evaluated in bins of $D(\x,\Psi)$. Their widths are generally comparable over the relevant range of \(D(\x,\Psi)\), although some differences are present.
}
}\label{fig:Lambda14}
\end{figure}

We first examine what unique information the tidal deformability constraint $\Lambda_{1.4}$ from GW170817 can provide on the central EOS-parameter $\x$ of canonical NSs.
In FIG.\,\ref{fig:Lambda14}, the correlation between $\Lambda_{1.4}$ and the scaling factor $D(\x,\Psi)$ is shown using $10^3$ randomly sampled EOSs for each of the nonlinear Walecka model and the meta-model EOS framework, yielding a total of $2\times10^3$ EOS samples. All EOSs are required to satisfy existing terrestrial and astrophysical constraints, as well as low-density nuclear-theory constraints. In these simulations, the corresponding maximum NS masses are filtered by the requirement $M_{\rm{NS}}^{\max}/M_\odot \gtrsim 2.01$.
For each EOS sample, the log-stability slope $\Psi$ is determined self-consistently from the corresponding $M_{\rm{NS}}$-$\varepsilon_{\rm c}$ relation. The resulting dependence of $\Psi$ on $M_{\rm{NS}}$ is shown in App.\,FIG.\,\ref{fig:psi} in the Appendix \ref{app-b}. Statistically, we have $\Psi_{1.4}\approx2.66\pm0.40$ for canonical NSs. The scaling factor $D(\x,\Psi)$ is then evaluated using Eqs.\,\eqref{eq:k2} and \eqref{eq:xi}, while $\Lambda_{1.4}$ is obtained directly from its definition. See Appendix \ref{app-a} for a detailed description of the EOS samples. Although a total of $2\times10^3$ EOS samples are included in the analysis, the scaling relation converges rapidly once about one to two hundred EOSs are used.
We note that the residual scatter is not uniform across the entire $D$ range, but increases moderately toward larger values of $D$. Accordingly, we employ a feasible-weighted-least-squares (FWLS) regression to account for the heteroscedastic residuals. The resulting {prediction interval (PI)} therefore depends on $D$, yielding a more accurate characterization of the local uncertainty than a constant-width confidence band.
Numerically, we find that:
\begin{equation}\label{eq:lambda14}
\mbox{for canonical NSs: }
\Lambda_{1.4}
\approx
6.61^{+0.03}_{-0.03}D(\x,\Psi)
+
8.35^{+1.55}_{-1.55},
\end{equation}
in the 68\% PI,
with a coefficient of determination $\mathrm{R}^2\approx0.960$; the corresponding results are summarized in TAB.\,\ref{tab:model_dependence}.
For simplicity, we omit the subscript ``1.4'' on the parameter $\x$.
Besides, 33 widely used microscopic EOSs from the literature are also shown in the plot (green ``+'' symbols). 
These EOSs span several major classes of existing models in the literature, including microscopic nucleonic calculations, relativistic and non-relativistic density-functional approaches, hyperonic and $\Delta$-admixed matter, as well as hybrid quark-hadron EOS constructions\,\cite{Typel15,Ofengeim24}. Although these EOSs are not included in the fitting procedure, they likewise cluster closely around the scaling relation extracted from the $2\times10^3$ calibrated EOS samples, exhibiting only mild diffusion, as illustrated in the figure. This agreement indicates that the combination of Eq.\,\eqref{eq:k2}, Eq.\,\eqref{eq:xi}, and Eq.\,\eqref{eq:sc2} effectively captures the essential physics underlying the behavior of $\Lambda$.
The remaining deviations are confined to only a few EOSs. A detailed examination shows that these outliers possess rather unusual physical properties. In particular, the three meta-model outliers correspond to exceptionally small symmetry energy slopes at saturation density, $L\approx22$--$28\,\rm{MeV}$, whereas the Walecka and microscopic EOS GNH3 outliers predict comparatively large radii of about $14.5\,\rm{km}$ for canonical NSs. These EOSs therefore occupy the extremes of the currently allowed parameter space and are not representative of the majority of physically viable EOSs. Consequently, they have little influence on the calibration of the scaling relation or on the inferred range of the central EOS-parameter $\x$.

\renewcommand*\figurename{\small FIG.}

\renewcommand*\tablename{\small TAB.}
\begin{table}[h!]
%\begin{ruledtabular}
\begin{tabular}{ccccc}
\\\hline
NSs & EOSs & $a$ & $b$ & $\rm{R}^2$ \\
\hline\hline
\multirow{3}{*}{$1.4M_\odot$}
& Walecka  & $6.57 \pm 0.04$ & $13.51 \pm 2.56$ & $0.956$ \\
& Meta     & $6.58 \pm 0.04$ & $9.05 \pm 2.43$ & $0.951$ \\
& combined & $6.61 \pm 0.03$ & $8.35 \pm 1.55$ & $0.960$ \\
\hline\hline
\multirow{3}{*}{TOV}
& Walecka  & $2.67 \pm 0.01$ & $1.97 \pm 0.02$ & $0.996$ \\
& Meta     & $2.72 \pm 0.03$ & $2.46 \pm 0.05$ & $0.836$ \\
& combined & $2.55 \pm 0.01$ & $2.68 \pm 0.04$ & $0.951$ \\
\hline
\end{tabular}
\caption{
Fitting coefficients of the scaling relation $\Lambda=aD+b$ for canonical and TOV NSs obtained from different EOS ensembles; $\mathrm{R}^2$ denotes the coefficient of determination. The quoted uncertainties represent the root-mean-square (RMS) scatter of the fitted coefficients evaluated over individual EOS samples.
}
\label{tab:model_dependence}
%\end{ruledtabular}
\end{table}

The scaling relation of Eq.\,\eqref{eq:lambda14} is particularly useful because it allows one to directly ``read off'' the central EOS-parameter $\x$ from the measured/observed tidal deformability. For instance, adopting $\Lambda_{1.4}\approx190$ from GW170817 yields $\x\approx0.182\pm0.002$, where the uncertainty in $\x$ arises mainly from the scaling relation itself. In this sense, the $\Lambda$-$D$ scaling permits a determination of $\x$ with an accuracy of about $1.1\%$, provided that $\Lambda$ is measured accurately.
We note that the inferred value is somewhat larger than conventional constraints. This discrepancy is partly attributable to the still considerable uncertainties in both the observationally measured and theoretically inferred values of $\Lambda$. For example, several recent studies\,\cite{Most18,Dong2025,Golomb2025} suggest that the most probable value of $\Lambda_{1.4}$ may be somewhat larger than 190.
Conversely, if one artificially adopts $\x\approx0.16$\,\cite{CaiLi2025Review}, the corresponding tidal deformability extracted from the $\Lambda$-$D$ scaling becomes $\Lambda_{1.4}\approx295\pm12$ with an uncertainty about $\pm4\%$.
Moreover, using the upper bound $\Lambda_{1.4}\approx580$ from GW170817, one obtains $\x\approx0.133\pm0.002$, corresponding to a relative uncertainty of about $1.5\%$. The absolute uncertainty in this case is at a comparable level to that obtained previously using $\Lambda_{1.4}\approx190$.
In the following, we adopt $\Lambda_{1.4}\approx190$ as the lower bound for canonical NSs. Consequently, the observational range $190\lesssim\Lambda_{1.4}\lesssim580$ implies
$0.133\lesssim\x\lesssim0.182$.

Interestingly, this range is consistent with the values independently inferred from the compactness scaling\,\cite{CaiLi2025Trace} or the joint mass-radius scalings\,\cite{CaiLiZhang2023ApJ}. The convergence of these distinct macroscopic observables toward a common range of $\x$ indicates that the global gravitational properties of NSs are controlled by a unified underlying core EOS-parameter. This agreement not only demonstrates the internal self-consistency of the IPAD-TOV approach, but also suggests that observables such as mass $M_{\rm{NS}}$, radius $R$, compactness $\xi$, and tidal deformability $\Lambda$ collectively encode coherent information about the microphysics of ultradense matter in NS cores.
A similar empirical correlation between $\x$ and $\Lambda$ was analyzed in Ref.\,\cite{Saes21}.

The central SSS relation of Eq.\,\eqref{eq:sc2}, together with $\Psi\approx2.66\pm0.40$ for canonical NSs (see Appendix \ref{app-b}), imposes an upper bound on the central EOS-parameter, $\x\lesssim0.25\pm0.01$, from the causality condition $s_{\rm c}^2\leq1$. Combined with the $\Lambda$-$D$ scaling shown in FIG.\,\ref{fig:Lambda14}, this leads to a causality lower bound on the tidal deformability of canonical NSs, $\Lambda_{1.4}\gtrsim69\pm7$, very close to the lower limit of about 70 inferred from GW170817\,\cite{Abbott2018}. Furthermore, once $\Lambda_{1.4}$ is observationally constrained, the central SSS can be inferred directly. For example, adopting $190\lesssim\Lambda_{1.4}\lesssim580$ together with the corresponding $\Psi$ values for canonical NSs, Eq.\,\eqref{eq:sc2} yields
$0.60_{-0.01}^{+0.01}\gtrsim s_{\rm c}^2 \gtrsim 0.39_{-0.01}^{+0.01}$. These results indicate that current GW observations are already probing the collective properties of matter in the innermost NS core, suggesting that the dense matter inside canonical NSs remains significantly away from the conformal limit characterized by $s^2\to1/3$ and $\phi\to1/3$\,\cite{Fuji22,JHChen24}.

\section{How close are NSs to the black-hole (BH) limit in dimensionless tidal deformability?}
\label{sec:AI-2}

\begin{figure}[h!]
\centering
\includegraphics[width=7.5cm]{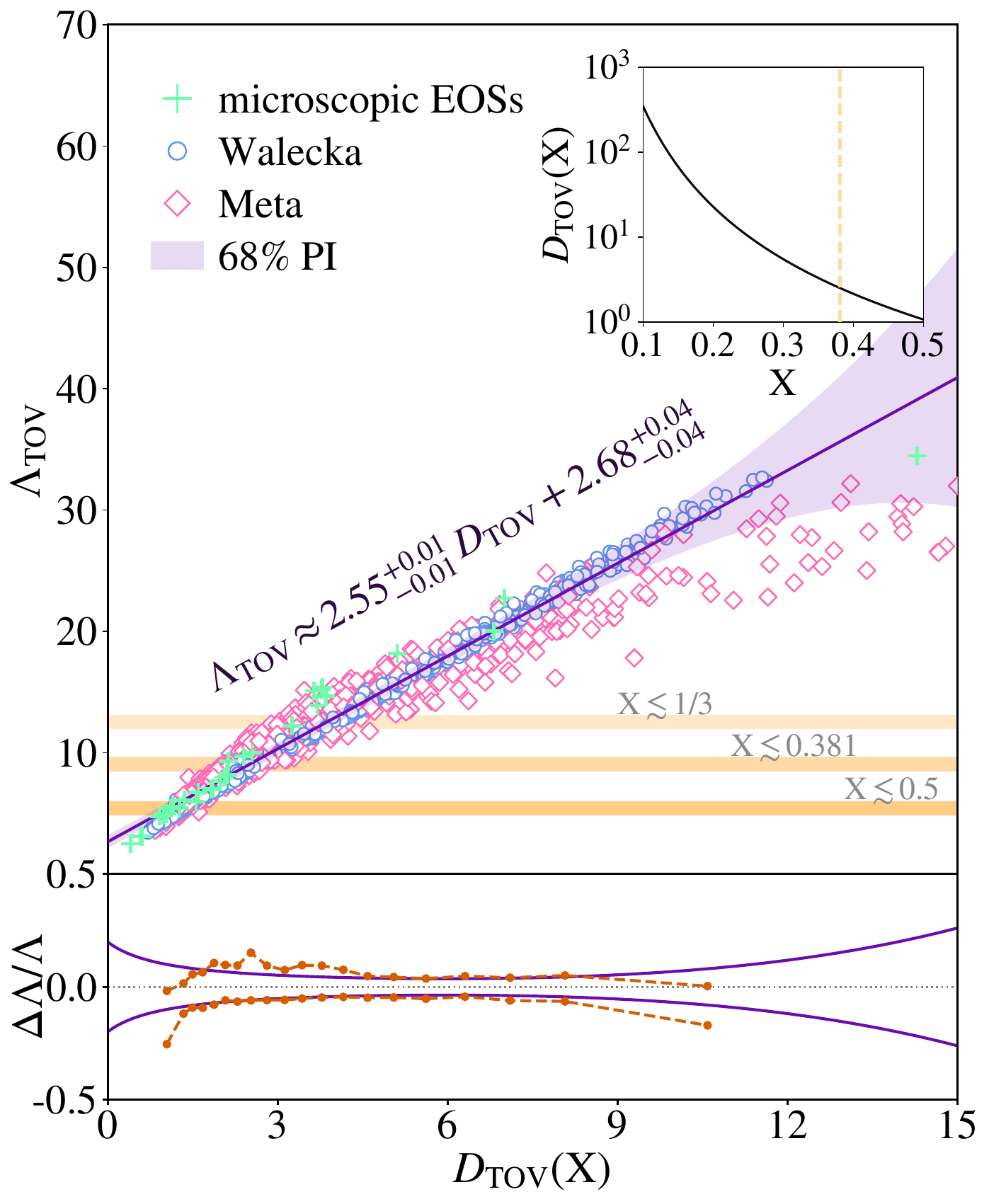}
\caption{(Color Online). The same as FIG.\,\ref{fig:Lambda14} but for correlation between $\Lambda_{\mathrm{TOV}}$ and $D_{\mathrm{TOV}}$. The light-yellow, orange, and orange-red bands (from top to bottom) correspond to the limits $\x \lesssim 1/3$ ($\Lambda_{\rm{TOV}}\gtrsim12.5_{-0.5}^{+0.5}$), $\x \lesssim 0.381$ ($\Lambda_{\rm{TOV}}\gtrsim9.1_{-0.5}^{+0.5}$), and $\x \lesssim 0.5$ ($\Lambda_{\rm{TOV}}\gtrsim5.4_{-0.5}^{+0.5}$), respectively. The inset illustrates the monotonicity of $D_{\mathrm{TOV}}(\x)$ over $\x$, where the dotted line indicates the limit $\x\lesssim0.381$.
{Similar to FIG.\,\ref{fig:Lambda14}, the lower panel compares the FWLS 68\% prediction interval (purple) with the empirical 16th-84th percentile interval (orange) evaluated in bins of $D_{\rm{TOV}}$; their widths are generally comparable in the region relevant to the inferred lower bound.
}
}\label{fig:LambdaTOV}
\end{figure}

Similar to the $\Lambda$-$D$ scaling for canonical NSs, one can also establish a corresponding relation for TOV NSs, for which $\Psi=0$. Using the $2\times10^3$ constructed EOS samples, we present the resulting $\Lambda$-$D$ correlation in FIG.\,\ref{fig:LambdaTOV}. In particular, we obtain
\begin{equation}\label{LambdaTOV}
\mbox{for TOV NSs: }
\Lambda_{\mathrm{TOV}}
\approx
2.55^{+0.01}_{-0.01} D_{\mathrm{TOV}}
+
2.68^{+0.04}_{-0.04},
\end{equation}
in the 68\% PI, where $D_{\mathrm{TOV}}\equiv D(\x,\Psi=0)$.
The linear regression gives a coefficient of determination $\mathrm{R}^2\approx0.951$ when the two EOS classes are combined, see TAB.\,\ref{tab:model_dependence}, which shows the strong tightness of scaling. The $(D_{\mathrm{TOV}},\Lambda_{\mathrm{TOV}})$ values for 33 additional microscopic EOSs are also displayed in FIG.\,\ref{fig:LambdaTOV}, and are found to approximately follow the relation of Eq.\,\eqref{LambdaTOV}.

While the scaling relation remains highly EOS-insensitive over the full ensemble, a mild model dependence emerges in the large-$D_{\rm{TOV}}$ region, where the Walecka and meta-model EOSs exhibit slightly different trends. This behavior likely reflects the increasing sensitivity of the tidal response to the detailed high-density microphysics in extremely compact configurations. The additional microscopic EOSs shown in FIG.\,\ref{fig:Lambda14} and FIG.\,\ref{fig:LambdaTOV} likewise exhibit slight deviations from the best-fit relations, particularly in the large-$D$ region, while remaining broadly consistent with the overall scaling. Nevertheless, these systematic deviations remain much smaller than the spread of the full EOS ensemble and are therefore not expected to qualitatively affect the EOS-insensitive character of the proposed scaling relation.

Because $D_{\mathrm{TOV}}$ is a monotonically decreasing function of $\x$ within the physically allowed range, imposing an upper bound on $\x$ immediately leads to a corresponding lower bound on $D_{\mathrm{TOV}}$, which in turn yields a lower bound on $\Lambda_{\mathrm{TOV}}$. For example, adopting the bound $\x\lesssim0.381$\,\cite{CaiLi2025Trace} and using the scaling relation in Eq.\,\eqref{LambdaTOV}, we obtain $
\Lambda_{\mathrm{TOV}} \gtrsim 9.1 \pm 0.5$.
Likewise, for typical TOV NSs with $\x\gtrsim0.15$, Eq.\,\eqref{eq:k2} yields approximately $0.02\lesssim k_2\lesssim0.04$.
Furthermore, TOV NSs contain the densest stable visible matter known in the Universe and therefore represent the closest observable configurations to the black hole (BH) limit. In this context, the tidal deformability $\Lambda$ provides a natural quantitative measure of the compactness and deformability of dense matter under extreme gravity. Smaller values of $\Lambda$ correspond to more compact and less deformable NSs, while the limiting case $\Lambda_{\rm{BH}}=0$ characterizes BHs. Therefore, the lower bound $\Lambda_{\mathrm{TOV}}\gtrsim9.1$ implies even the most compact stable NSs remain measurably deformable and distinct from BHs. This result provides a direct quantitative characterization of how closely the densest stable matter in the Universe approaches the BH limit.
Owing to the FWLS regression adopted in this work, the uncertainty associated with the inferred lower bound on $\Lambda_{\rm{TOV}}$ is evaluated locally near $D_{\rm{TOV}}$, rather than from a global residual estimate. The relatively small residual scatter in this low-$D_{\rm{TOV}}$ region therefore leads to a correspondingly tighter constraint on the inferred lower bound. To further assess the sensitivity of the inferred lower bound to the fitting procedure, we repeated the analysis using different fitting windows in the low-$D_{\rm{TOV}}$ region. Since $\x\approx0.381$ corresponds to approximately the lowest 37\% of the $D_{\rm{TOV}}$ distribution, we considered fitting windows covering the lowest 30\%, lowest 40\%, and the full $D_{\rm{TOV}}$ range. These yield $\Lambda_{\rm{TOV}}\gtrsim9.1\pm0.6$, $\Lambda_{\rm{TOV}}\gtrsim9.0\pm0.6$, and $\Lambda_{\rm{TOV}}\gtrsim9.1\pm0.5$, respectively. The close agreement among these results indicates that the inferred lower bound is robust against the choice of fitting window.
{As an additional check of the uncertainty estimate, the lower panels of FIGs.\,\ref{fig:Lambda14} and \ref{fig:LambdaTOV} compare the 68\% FWLS prediction intervals (purple) with the empirical 16th-84th percentile intervals (orange) obtained in bins of $D$ and $D_{\rm{TOV}}$, respectively. Although some differences are present, particularly for the TOV configurations, the empirical intervals are generally comparable to the FWLS prediction over the respective relevant ranges of $D$ or $D_{\rm{TOV}}$, providing an additional check on the local uncertainty estimates.
}

The upper bound $\x\lesssim0.381$ adopted in the present work was obtained within the IPAD-TOV framework\,\cite{CaiLi2025Trace}. It should therefore be regarded as an effective estimate rather than a universal physical limit. Owing to the highly nonlinear nature of the TOV equations, higher-order perturbative corrections may  modify this value. Likewise, nonperturbative features of the EOS, such as strong first-order phase transitions, may also affect the corresponding upper bound on $\x$. Nevertheless, the $\Lambda_{\mathrm{TOV}}$-$D_{\mathrm{TOV}}$ scaling established here remains applicable, and any adopted upper bound on $\x$ can be directly translated into a corresponding lower bound on $\Lambda_{\mathrm{TOV}}$.
Extensive studies have been carried out on the upper bound of $\x$ in NSs\,\cite{Cai2024Front}. To illustrate the dependence of the inferred lower bound on the adopted value of $\x$, we consider two additional cases. In the first, we adopt a relatively loose upper bound, $\x\lesssim0.5$, which encompasses a broader class of existing EOSs, while in the second we impose $\x\lesssim1/3$, motivated by the conformal limit\,\cite{Ann23}. These choices lead to $\Lambda_{\mathrm{TOV}}\gtrsim5.4\pm0.5$ and $\Lambda_{\mathrm{TOV}}\gtrsim12.5\pm0.5$, respectively, as illustrated in FIG.\,\ref{fig:LambdaTOV}. We note that several microscopic EOSs shown in the figure lie below the lower bound obtained for $\x\lesssim0.381$; however, all of these EOSs satisfy $\x\gtrsim0.381$ and therefore fall outside the adopted range used to derive this particular bound. 
{We further examine the microscopic EOSs with $\Lambda_{\rm{TOV}}<9.1$. These EOSs typically yield relatively large central squared sound speeds, $s_{\rm{c,TOV}}^2\approx0.64$-1.00 (with a median of 0.79), and highly compact TOV configurations, with compactnesses in the range 0.29-0.32. For a $2\,M_{\odot}$ NS, the corresponding radii range from approximately 9.2\,km to 10.2\,km, which is relatively small compared with current empirical radius estimates.}
Although the numerical values of the lower bounds vary with the adopted upper limit on $\x$, they remain mutually consistent and collectively demonstrate the robustness of the present framework for inferring lower bounds on $\Lambda_{\mathrm{TOV}}$.

\section{Jointly constraining the central EOS of canonical NSs by combining the mass and tidal-deformability scalings}
\label{AI-3}

\begin{figure}[h!]
\centering
\includegraphics[width=8.cm]{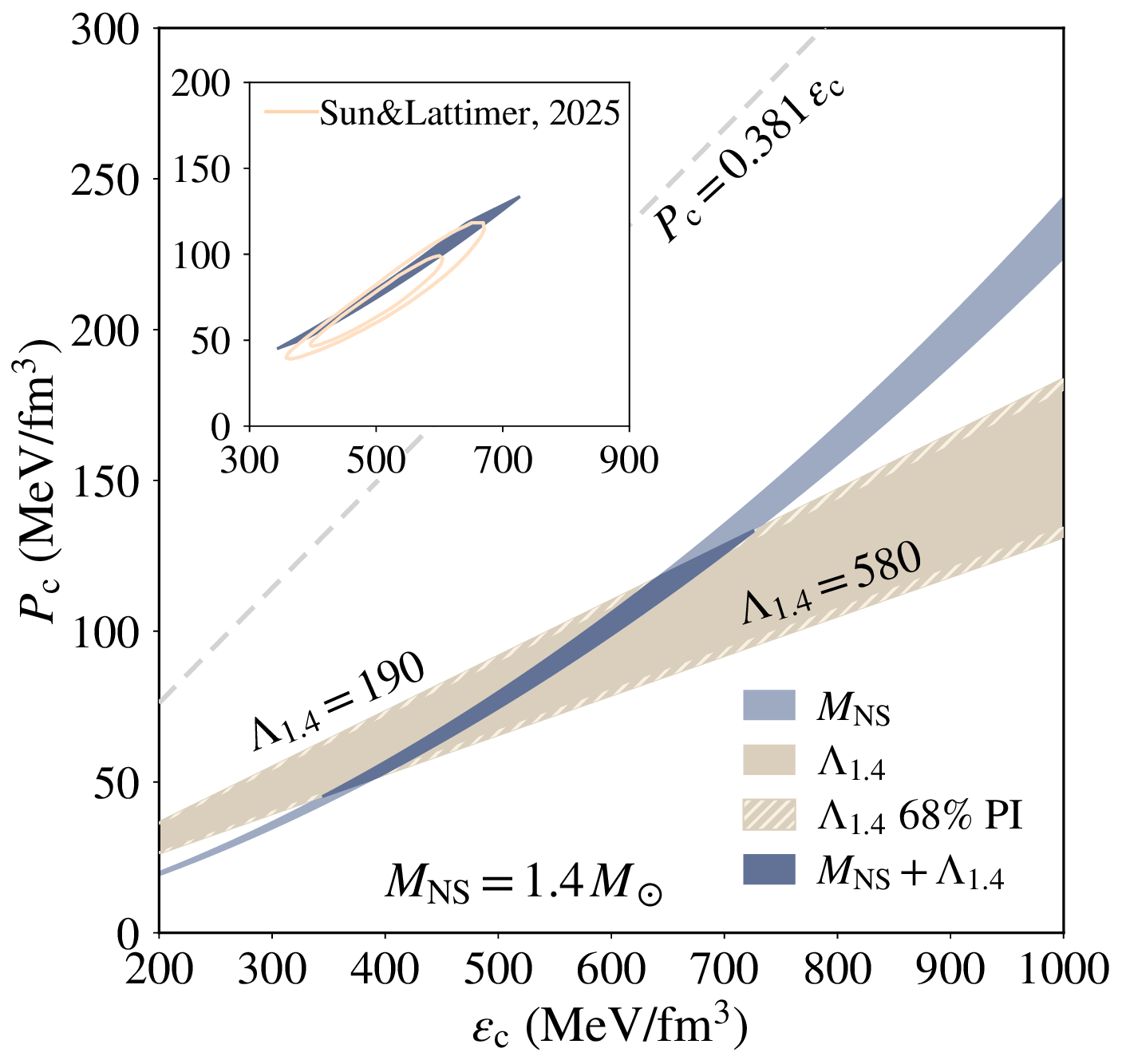}
\caption{(Color Online). Joint constraints on the central EOS $P_{\rm c}(\varepsilon_{\rm c})$ for canonical NSs from the mass scaling (deep-blue band) and the tidal-deformability scaling (beige band for $190\lesssim\Lambda_{1.4}\lesssim580$). The hatched regions corresponding to $\Lambda\approx190$ and $\Lambda\approx580$ indicate the associated 68\% PIs. 
{The effective upper bound }$P_{\rm{c}}/\varepsilon_{\rm{c}}\lesssim0.381$ is shown using the gray dashed line.
The inset compares our constraint on the central EOS with the recent analysis of Ref.\,\cite{Sun2025}, where the ellipses denote the 68\% and 95\% confidence regions. See the text for details.
}
\label{fig:constraint1}
\end{figure}

Since the tidal deformability is a dimensionless quantity, the $\Lambda$-$D$ scaling can only be used to extract the EOS-parameter $\x$. To further determine the EOS itself (the dependence of $P_{\rm c}$ on $\varepsilon_{\rm c}$), one should additionally employ scalings that depend separately on $P_{\rm c}$ and $\varepsilon_{\rm c}$, rather than only through the ratio $\x$.
Here, we combine the mass scaling with the $\Lambda$-$D$ scaling to achieve this goal. The mass scaling is based on the correlation between $M_{\rm{NS}}$ and $\Gamma_{\rm c}$, where $\Gamma_{\rm c}
=
\varepsilon_{\rm c}^{-1/2}
[\x/(1+3\x^2+4\x)]^{3/2}
[1+36\x(1+\Psi)/(50+37\Psi+5\Psi^2)]$, and the correction factor is identical to that appearing in Eq.\,\eqref{eq:xi}. This scaling is verified using our EOS ensembles in FIG.\,\ref{fig:Lambda14}, yielding a $1\sigma$ (68\%) regression half-width of about $3.2\%$.
The resulting joint constraint on the central EOS $P_{\rm c}(\varepsilon_{\rm c})$ is shown in FIG.\,\ref{fig:constraint1}.

By combining the mass-scaling with the tidal-deformability scaling and adopting $580\gtrsim\Lambda_{1.4}\gtrsim190$, one effectively obtains $373^{+37}_{-28}\,\rm{MeV}/\rm{fm}^3
\lesssim\varepsilon_{\rm c}\lesssim664^{+61}_{-44}\,\rm{MeV}/\rm{fm}^3
$ for the central energy density as well as $49.6^{+5.7}_{-4.3}\,\rm{MeV}/\rm{fm}^3\lesssim
P_{\rm c}\lesssim 120.7^{+12.7}_{-9.2}\,\rm{MeV}/\rm{fm}^3
$ for the central pressure, where the uncertainties attached to each bound originate from the $\Lambda$-$D$ scaling. One can also see from the figure that, owing to the FWLS analysis, the uncertainties associated with the $\Lambda$-$D$ scaling at $\Lambda_{1.4}\approx580$ and $\Lambda_{1.4}\approx190$ are at a comparable level, as indicated by the hatched regions corresponding to the 68\% PIs.
Introducing the energy density at nuclear saturation density, $\varepsilon_0\approx150\,\rm{MeV}/\rm{fm}^3$ for $\rho_0\approx0.16\,\rm{fm}^{-3}$, the above constraint for energy density may alternatively be written as
$ 2.49^{+0.25}_{-0.19} \lesssim \varepsilon_{\rm c}/\varepsilon_0 \lesssim 4.43^{+0.41}_{-0.29}
$. In particular, the upper bound on $\Lambda_{1.4}$ from GW170817 implies a lower bound on the reduced central energy density, namely $\varepsilon_{\rm c}/\varepsilon_0\gtrsim2.49$.
In the inset of FIG.\,\ref{fig:constraint1}, we compare the central EOS obtained in this work for canonical NSs with the recent analysis of Ref.\,\cite{Sun2025}, where a power-law fitting was adopted. One can clearly see that our constraint on the core EOS is generally consistent with that of Ref.\,\cite{Sun2025}, while exhibiting a narrower uncertainty band.

It is worth noting that the above constraints are obtained by combining the mass scaling and the tidal-deformability scaling, which are derived within the same IPAD-TOV framework and therefore are not strictly statistically independent. Nevertheless, they are based on different observables and governing equations. A Pearson correlation analysis yields a coefficient of determination of about $0.34$, indicating a moderate degree of shared linear correlation while leaving substantial complementary information in the two scalings. Their combination therefore provides tighter constraints on the central EOS than either scaling alone.

Based on the $\varepsilon_{\rm c}$ and $P_{\rm c}$ extracted here together with the inferred pressure around two times nuclear saturation density  of Ref.\,\cite{Abbott2018}, one may estimate the SSS profile of dense matter in canonical NSs. In particular, an average supranuclear stiffness $s^2\sim0.3$-$0.4$ is obtained between $\varepsilon\approx2\varepsilon_0$ and the central density region $\varepsilon_{\rm c}$. Since this interval-averaged value is already comparable to, and may even mildly exceed, the conformal limit $s^2=1/3$\,\cite{Ann23}, the local SSS inside canonical NSs likely approaches or temporarily violates the conformal bound above about $2\rho_0$, indicating a significant stiffening of dense matter in the NS core.

Our $\Lambda$-$D$ scaling implies that an upper bound on $\Lambda$ translates directly into a lower bound on $\x$. 
This lower bound on $\x$ in turn yields a lower bound on the compactness $\xi$\,\cite{CaiLi2025Trace}, which consequently leads to an upper bound on the stellar radius for a given mass. 
Using the EOS samples and the scaling relations, we thus infer
$
R_{1.4}\lesssim13.2_{-0.7}^{+0.7}\,\rm{km}$ from $\Lambda_{1.4}\lesssim580$.
This result is consistent with the upper bound for $R_{1.4}$ inferred from GW170817\,\cite{Abbott2018}, which was obtained mainly from the empirical correlation between $\Lambda_{1.4}$ and $R_{1.4}$. For completeness, we present the corresponding $\Lambda_{1.4}$-$R_{1.4}$ correlation extracted from our EOS samples in Appendix \ref{app-c} and App.\,FIG.\,\ref{fig:LambdaR}.

The robust scaling behavior emerging from the IPAD-TOV framework is remarkable, since the underlying derivations rely only on low-order expansions of the dimensionless stellar-structure and tidal-response equations, yet remain accurate across a broad range of realistic EOSs. This may indicate that the dimensionless TOV system possesses an intrinsic universality structure, in which much of the microscopic EOS dependence is effectively compressed into only a few macroscopic dimensionless quantities, such as the central EOS-parameter $\x$ and the log-stability slope $\Psi$. Meanwhile, the $\x=P_{\rm c}/\varepsilon_{\rm c}$ itself remains bounded by $\x\lesssim0.381$ for causal NS matter, which may further suppress higher-order corrections and improve the convergence of the expansion. Physically, the robustness of these scalings suggests that the equilibrium and tidal properties of NSs are governed primarily by global geometric and stability conditions rather than detailed microscopic features of the EOS, since the TOV equations effectively integrate the microphysics over the entire stellar configuration while suppressing higher-order EOS-dependent structures. Such behavior is qualitatively reminiscent of universality phenomena encountered in statistical physics and in the I-Love-Q relations\,\cite{YagiYunes2013a,YagiYunes2013b} of compact stars. It would therefore be interesting to further explore whether the IPAD-TOV framework reflects a deeper similarity or emergent structure within the dimensionless relativistic stellar equations.

\section{Summary}
\label{sec:summary}

In summary, we have uncovered a previously unrecognized scaling relation between the dimensionless tidal deformability $\Lambda$ and the central EOS parameter
$
\x \equiv {P_{\rm c}}/{\varepsilon_{\rm c}},
$
by analyzing the dimensionless stellar-structure and relativistic tidal-response equations within the IPAD-TOV framework. The resulting intrinsic tidal-response function $D(\x,\Psi)$ depends primarily on the central EOS parameter $\x$ and the logarithmic stability slope $\Psi$, and encapsulates the essential EOS dependence of $\Lambda$. The scaling relation was established analytically and validated using broad ensembles of calibrated EOSs constructed within both the nonlinear Walecka model and meta-model frameworks, as well as 33 representative microscopic EOSs from the literature. Despite the diversity of the underlying EOSs, including models featuring hyperons, quarks, smooth crossovers, and phase transitions, the $\Lambda$-$D$ correlation remains very tight, demonstrating its largely EOS-insensitive character.

For canonical $1.4 M_\odot$ NSs, the scaling enables a direct determination of the central EOS parameter $\x$ from the observed tidal deformability. Applying the observational range inferred from GW170817, $190 \lesssim \Lambda_{1.4} \lesssim 580$, yields
$
0.133 \lesssim \x \lesssim 0.182,
$
indicating that current inspiral GW observations already constrain the thermodynamic properties of matter deep inside NS cores. Combined with the previously established mass scaling, the new relation further constrains the central energy density and pressure of canonical NSs and provides insight into the corresponding sound-speed profile.

For maximum-mass stable configurations, the scaling predicts a robust lower bound on the tidal deformability,
$
\Lambda_{\rm{TOV}} \gtrsim 9.1^{+0.5}_{-0.5},
$
adopting {the effective limit $\x \lesssim 0.381$ from IPAD-TOV framework}. This result quantitatively demonstrates that even the most compact stable NSs remain measurably deformable and distinctly separated from black holes, for which $\Lambda_{\rm{BH}}=0$.

The emergence of this scaling suggests that much of the microscopic EOS dependence encoded in the relativistic stellar-structure and tidal-response equations can be effectively compressed into a small set of dimensionless macroscopic quantities, primarily $\x$ and $\Psi$. In this sense, tidal deformability is not merely a probe of matter near twice nuclear saturation density, but also carries direct information about the EOS of the stellar core.

Looking ahead, future high-precision gravitational-wave observations, together with improved measurements of NS masses and radii, will enable increasingly precise determinations of $\x$, the central pressure, and the central energy density. The scaling relation identified here thus establishes a new and largely model-insensitive framework for connecting gravitational-wave observables directly to the microphysics of ultradense matter in NS interiors.

\section*{Acknowledgment} 
We would like to thank Ying Zhou for helpful discussions.
This work was supported in part by the National Natural Science Foundation of China under contract No. 12547102, the U.S. Department of Energy, Office of Science, under Award Number DE-SC0013702, the CUSTIPEN (China-U.S. Theory Institute for Physics with Exotic Nuclei) under the US Department of Energy Grant No. DE-SC0009971.\\

\appendix
\section*{Appendices}

In the following, we provide detailed information on several technical aspects of the main text. Appendix \ref{app-a} introduces the basic definitions of the EOS for asymmetric nuclear matter (ANM). Appendix \ref{app-b} is devoted to the description of the $\beta$-stable EOS used in this work.  Appendix \ref{app:A} presents the analytical expressions for the key quantities characterizing the tidal deformability.
Finally, Appendix \ref{app-c} gives the correlation between $\Lambda_{1.4}$ and $R_{1.4}$ using our EOS ensemble.

\setcounter{figure}{0}

\section{Basic Definitions on EOS of ANM}\label{app-a}

The EOS of isospin asymmetric matter, defined by its binding energy per nucleon, can be expressed as\,\cite{LCK08}
\begin{equation}
  E(\rho, \delta) \approx E_0(\rho) + E_{\text{sym}}(\rho)\delta^2 + \mathcal{O}(\delta^4),
  \label{eq:e-expansion}
\end{equation}
where $\rho = \rho_\rm n+\rho_\rm p$ is the nucleon density with $\rho_\rm n$ ($\rho_\rm p$) denoting the neutron (proton) density, $\delta = (\rho_\rm n - \rho_\rm p)/(\rho_\rm n + \rho_\rm p)$ is the isospin asymmetry between n and p, $E_0(\rho) \equiv E(\rho, \delta = 0)$ is the EOS of symmetric nuclear matter (SNM), and the symmetry energy $E_{\mathrm{sym}}(\rho)$ is defined as\,\cite{LCK08,LCXZ21,AnR24,Ding24,LiuYY25}
\begin{equation}
  E_{\mathrm{sym}}(\rho) \equiv \frac{1}{2} \frac{\partial^2 E(\rho, \delta)}{\partial \delta^2} \bigg|_{\delta=0}.
\end{equation}
At nuclear matter saturation density $\rho_0$, $E_0(\rho)$ can be expanded in $\chi \equiv (\rho - \rho_0)/3\rho_0$ as
\begin{equation}
  E_0(\rho) \approx E_0(\rho_0) + \frac{K_0}{2!}\chi^2 + \frac{J_0}{3!}\chi^3 + \frac{I_0}{4!}\chi^4 + \mathcal{O}(\chi^5),
  \label{eq:e0-expansion}
\end{equation}
where $E_0(\rho_0)$ is the binding energy per nucleon in SNM at $\rho_0$, $K_0$ is the incompressibility coefficient\,\cite{Wang25}, $J_0$ is the skewness coefficient\,\cite{Cai17J0}, and $I_0$ is the kurtosis coefficient. Similarly, $E_{\mathrm{sym}}(\rho)$ can be expanded as
\begin{equation}
  E_{\text{sym}}(\rho) \approx E_{\text{sym}}(\rho_0) + L\chi + \frac{K_\mathrm{sym}}{2!} \chi^2 + \frac{J_\mathrm{sym}}{3!}\chi^3 + \mathcal{O}(\chi^4),
  \label{eq:esym-expansion}
\end{equation}
where $L$ is the slope of the symmetry energy at $\rho_0$, $K_\mathrm{sym}$ and $J_{\rm{sym}}$ are the curvature and skewness coefficients of the symmetry energy\,\cite{LCXZ21}, respectively. 

\section{EOS for NS Matter}\label{app-b}

In this appendix, we describe the $\beta$-stable and charge neural NS EOS employed in this work. We first introduce two core EOS frameworks, namely the nonlinear Walecka model in the mean-field approximation and the meta-model, and then present the construction of the complete NS EOS. We also provide numerical results for the NS mass-radius (M-R) relations, together with the corresponding log-stability slope $\Psi=2\d\ln M_{\rm{NS}}/\d\ln\varepsilon_{\rm c}$.

\subsection{Nonlinear Walecka Model}

Firstly, we introduce the nonlinear Walecka model under the mean field approximation\,\cite{SW1986,Mueller1996}.
The Lagrangian of this model reads\,\cite{Li2022}:
\begin{align}
    \mathcal{L} = &\overline{\psi}\left(i\partial_\mu\gamma^\mu - M_{\rm N}\right)\psi \notag\\
    &+ g_\sigma\sigma\overline{\psi}\psi - g_\omega\omega_\mu\overline{\psi}\gamma^\mu\psi 
    - g_\rho\vec{\rho}_\mu\overline{\psi}\gamma^\mu\vec{\tau}\psi 
    + g_\delta\vec{\delta}\overline{\psi}\vec{\tau}\psi \notag\\
    &+ \frac{1}{2}\left(\partial_\mu\sigma\partial^\mu\sigma - m_\sigma^2\sigma^2\right)
    - \frac{1}{3}b_\sigma M_{\rm N}(g_\sigma\sigma)^3
    - \frac{1}{4}c_\sigma(g_\sigma\sigma)^4 \notag\\
    &- \frac{1}{4}\omega_{\mu\nu}\omega^{\mu\nu}
    + \frac{1}{2}m_\omega^2\omega_\mu\omega^\mu
    + \frac{1}{4}c_\omega(g_\omega^2\omega_\mu\omega^\mu)^2\notag \\
    &- \frac{1}{4}\vec{\rho}_{\mu\nu}\vec{\rho}^{\mu\nu}
    + \frac{1}{2}m_\rho^2\vec{\rho}_\mu\vec{\rho}^\mu
    + \frac{1}{2}\Lambda_{\rm V}g_\rho^2\vec{\rho}_\mu\vec{\rho}^\mu g_\omega^2\omega_\mu\omega^\mu\notag \\
    &+ \frac{1}{2}\left(\partial_\mu\vec{\delta}\partial^\mu\vec{\delta} 
    - m_\delta^2\vec{\delta}^2\right)
    + \frac{1}{2}C_{\delta\sigma}g_\sigma^2\sigma^2g_\delta^2\vec{\delta}^2, \label{eq:20}
\end{align}
where $M_{\rm N}\approx939\,\rm{MeV}$ denotes the static nucleon mass, while $m_\sigma$, $m_\omega$, $m_\rho$, and $m_\delta$ are the meson masses. The quantities $g_\sigma$, $g_\omega$, $g_\rho$, and $g_\delta$ correspond to the coupling constants between nucleons and the associated mesons. In addition, $b_\sigma$ and $c_\sigma$ characterize the self-interaction of the $\sigma$ meson, $c_\omega$ describes the self-interaction of the $\omega$ meson\,\cite{Mueller1996}, $\Lambda_{\rm V}$ represents the coupling between the $\vec{\rho}$ and $\omega$ mesons\,\cite{ToddRutel2005}, and $C_{\delta\sigma}$ denotes the coupling between the $\sigma$ and $\vec{\delta}$ mesons\,\cite{Li2022}. 

The EOS of ANM obtained from the Lagrangian \eqref{eq:20} has been successfully applied to studies of low-density nuclear structure, intermediate-energy heavy-ion collisions, and high-density NS matter\,\cite{Li2022}. It therefore provides a suitable and well tested framework for extracting the $\Lambda$-scaling in the present work.

From the Lagrangian density $\mathcal{L}$, one can derive the equations of motion for the meson fields. Within the mean-field approximation, fluctuations and correlations are neglected, and the meson fields in the resulting equations are replaced by their expectation values\,\cite{Li2022}:
\begin{align}
    m_\sigma^2\overline{\sigma} &= g_\sigma\bigg[\rho_\rm p^{\rm s} + \rho_\rm n^{\rm s} - b_\sigma M_{\rm N}(g_\sigma\overline{\sigma})^2 \notag\\
        &\hspace{1cm} - c_\sigma (g_\sigma\overline{\sigma})^3 
     + C_{\delta\sigma}g_\sigma\overline{\sigma}\left(g_\delta\overline{\delta}^{(3)}\right)^2\bigg], 
        \label{eq:21} \\
    m_\omega^2\overline{\omega}_0 &= g_\omega\left[\rho_{\rm n}+\rho_{\rm p} - c_\omega\left(g_\omega\overline{\omega}_0\right)^3 
        - \Lambda_{\rm V}g_\omega\overline{\omega}_0\left(g_\rho\overline{\rho}_0^{(3)}\right)^2\right], 
        \label{eq:22} \\
    m_\rho^2\overline{\rho}_0^{(3)} &= g_\rho\left[\rho_\rm p - \rho_\rm n
        - \Lambda_{\rm V} g_\rho\overline{\rho}_0^{(3)}\left(g_\omega\overline{\omega}_0\right)^2\right], 
        \label{eq:23} \\
    m_\delta^2\overline{\delta}^{(3)} &= g_\delta\left[\rho_\rm p^{\rm s} - \rho_\rm n^{\rm s} 
        + C_{\delta\sigma} g_\delta\overline{\delta}^{(3)}\left(g_\sigma\overline{\sigma}\right)^2\right].
        \label{eq:24}
\end{align}
Here, the overline indicates the expectation value of the corresponding meson field. The subscript ``0'' labels the zeroth component of the four-vector, while the superscript ``(3)'' denotes the third component in isospin space. 
The scalar density $\rho_J^{\rm s}$ is defined as\,\cite{Cai2014,WeiSN24}
\begin{align}
    \rho_J^{\rm s} &= \frac{2}{(2\pi)^3} \int_0^{k_{\rm F}^J} \frac{M_J^*}{\sqrt{\v{k}^2 + M_J^{*,2}}} \d\v{k} \notag\\
    &= \frac{M_J^*}{2\pi^2} \left[ k_\rm F^J  \varepsilon_\rm F^{*,J} - M_J^{*,2} \ln \left(\frac{k_\rm F^J + \varepsilon_\rm F^{*,J}}{M_J^*}\right) \right], \quad J = \rm n, \rm p.
\end{align}
In the above expression, $\varepsilon_\rm F^{*, J} = \sqrt{k_\rm F^{J,2} + {M_J^{*,2}}}$, where $M_J^*$ denotes the effective (Dirac) mass of the neutron or proton\,\cite{Cai2014,Qin25},
\begin{equation}
    M_J^* = M_{\rm N} - g_\sigma \overline{\sigma} - g_\delta \overline{\delta}^{(3)} \tau_3^J, \quad J = \rm n, \rm p, 
\end{equation}
and $k_{\rm F}^J = k_{\rm F}(1 + \tau_3^J \delta)^{1/3}$ is the corresponding Fermi momentum, with $\tau_3^J = +1(-1)$ for neutrons (protons), while $k_\rm F$ denotes the Fermi momentum of SNM.

Within the mean-field approximation, the energy density $\varepsilon = \langle \mathcal{T}^{00}\rangle$ and the pressure $P=3^{-1}\sum_{j=1}^3\langle\mathcal{T}^{jj}\rangle$ can be derived from the energy-momentum tensor $\mathcal{T}^{\mu \nu}$ associated with the interacting Lagrangian density in Eq.\,\eqref{eq:20}. The resulting expressions for this nuclear matter system are\,\cite{Li2022}
\begin{align}
    \varepsilon =  & \varepsilon_\rm p^{\mathrm{kin}} + \varepsilon_\rm n^{\mathrm{kin}} \notag \\
    &+ \frac{1}{2} m_\sigma^2 \overline{\sigma}^2 + \frac{1}{2} m_\omega^2 \overline{\omega}_0^2 
    + \frac{1}{2} m_\rho^2 \overline{\rho}_0^{(3),2} + \frac{1}{2} m_\delta^2 \overline{\delta}^{(3),2} \notag \\
    &+ \frac{1}{3} M_{\rm N} b_\sigma (g_\sigma \overline{\sigma})^3 + \frac{1}{4} c_\sigma (g_\sigma \overline{\sigma})^4 + \frac{3}{4} c_\omega \left(g_\omega \overline{\omega}_0\right)^4 \notag \\
    &+ \frac{3}{2} \Lambda_{\rm V} \left(g_\rho \overline{\rho}_0^{(3)}\right)^2 \left(g_\omega \overline{\omega}_0\right)^2 - \frac{1}{2} C_{\delta\sigma} \left(g_\delta \overline{\delta}^{(3)}\right)^2 (g_\sigma \overline{\sigma})^2, \label{eq:27}
\end{align}
and\,\cite{Li2022}
\begin{align}
P =& P_\rm p^{\mathrm{kin}} + P_\rm n^{\mathrm{kin}} \notag \\
&- \frac{1}{2} m_\sigma^2 \overline{\sigma}^2 + \frac{1}{2} m_\omega^2 \overline{\omega}_0^2 + \frac{1}{2} m_\rho^2 \overline{\rho}_0^{(3),2} - \frac{1}{2} m_\delta^2 \overline{\delta}^{(3),2} \notag \\
&- \frac{1}{3} M_{\rm N} b_\sigma (g_\sigma \overline{\sigma})^3 - \frac{1}{4} c_\sigma (g_\sigma \overline{\sigma})^4 + \frac{1}{4} c_\omega \left(g_\omega \overline{\omega}_0\right)^4 \notag \\
&+ \frac{1}{2} \Lambda_{\rm V} \left(g_\rho \overline{\rho}_0^{(3)}\right)^2 \left(g_\omega \overline{\omega}_0\right)^2 + \frac{1}{2} C_{\delta\sigma} \left(g_\delta \overline{\delta}^{(3)}\right)^2 (g_\sigma \overline{\sigma})^2, \label{eq:28}
\end{align}
where $\varepsilon_J^{\mathrm{kin}}$ and $P_J^{\mathrm{kin}}$ are given by
\begin{align}
\varepsilon_J^{\mathrm{kin}} &= \frac{2}{(2\pi)^3} \int_0^{k_\rm {\rm F}^J}\d\v{k} \sqrt{\v{k}^2 + M_J^{*,2}},\\
P_J^{\mathrm{kin}} &= \frac{1}{3}\frac{2}{(2\pi)^3} \int_0^{k_\rm F^J} \d\v{k}\frac{\v{k}^2}{\sqrt{\v{k}^2 + M_J^{*,2}}} .
\end{align}
The EOS of ANM is then obtained via:
\begin{equation}
E(\rho,\delta)=\varepsilon(\rho,\delta)/\rho-M_{\rm N}.\end{equation}
Using these equations, one can further derive analytical expressions for various related physical quantities, including the nucleon Dirac mass $M_0^*(\rho_0)$ in SNM, the binding energy $E_0(\rho_0)$ per nucleon in SNM, the incompressibility coefficient $K_0$ of SNM, the skewness coefficient $J_0$ of SNM, the symmetry energy $E_\mathrm{sym}(\rho_0)$ as well as the slope $L$ of the symmetry energy; see Ref.\,\cite{Li2022,Cai2014} for additional details.

\subsection{Meta-model EOSs}

The meta-model provides a flexible and largely model-independent framework for constructing the EOS of dense nuclear matter. Instead of adopting a specific microscopic interaction or energy-density functional, it characterizes the EOS directly in terms of the empirical nuclear-matter parameters introduced in Eqs.~\eqref{eq:e-expansion}--\eqref{eq:esym-expansion}. In this approach, the EOS of symmetric nuclear matter (SNM), $E_0(\rho)$, and the symmetry energy, $E_{\mathrm{sym}}(\rho)$, are expanded around the saturation density $\rho_0$ in powers of $\chi=(\rho-\rho_0)/3\rho_0$, with the EOS determined by parameters such as $K_0$, $J_0$, $E_{\rm{sym}}(\rho_0)$, $L$, $K_{\mathrm{sym}}$, and $J_{\mathrm{sym}}$, etc. In Bayesian analyses, they are just parameterizations with the last parameters containing all information at high densities. Unlike most of the microscopic interaction parameters, there are empirical ranges for these parameters, largely based on experiments, to be used in setting their prior ranges in Bayesian analyses. By expressing the EOS directly in terms of these empirical quantities, the meta-model can reproduce the predictions of a broad class of relativistic and non-relativistic nuclear many-body approaches within a unified framework\,\cite{ZhangLi2018,ZhangLi2019,XieLi2020,ZhangLi2021}.

A major advantage of the meta-model is that it separates the generic thermodynamic properties of dense matter from the details of specific nuclear interactions. This feature makes it particularly suitable for systematic uncertainty quantification and correlation studies involving dense matter. Moreover, the resulting EOS ensemble spans a much broader range of behaviors than most conventional microscopic models, including some extreme or even potentially unphysical cases. Such a broad exploration of the EOS parameter space is especially valuable for identifying intrinsic and robust scaling relations\,\cite{CaiLi2025Trace} that are insensitive to the details of the nuclear interaction or the underlying many-body framework.

The meta-model is also particularly well-suited for Bayesian inference of NS EOS parameters. Since the TOV equations depend only on the macroscopic relation $P(\varepsilon)$ and not on its microscopic origin, NS global observables are intrinsically composition-degenerate. Consequently, the inverse mapping from observed stellar properties to a unique microscopic Hamiltonian is generally impossible. Instead, NS masses, radii, and tidal deformabilities constrain the EOS $P(\varepsilon)$ and its derivatives, rather than the underlying microscopic model itself\,\cite{Li26}. In this regard, the meta-model provides a natural intermediate representation between microscopic nuclear theories and macroscopic NS observables. It retains the essential thermodynamic information encoded in the EOS while remaining largely agnostic about the detailed microscopic dynamics, making it particularly effective for extracting robust EOS information from astrophysical observations.

\subsection{$\beta$-stable NS EOS and sampling schemes}

Here, we describe the construction of the charge neutral and $\beta$-stable NS EOSs adopted in this work, together with the corresponding sampling schemes for the empirical nuclear matter parameters. NSs generally consist of three regions: the core, the inner crust and the outer crust. We employ either the nonlinear Walecka model or the meta-model to construct the EOS of the NS core.

\renewcommand*\figurename{\small App.\,FIG.}
\begin{figure}[h!]
\centering
\includegraphics[width=7.5cm]{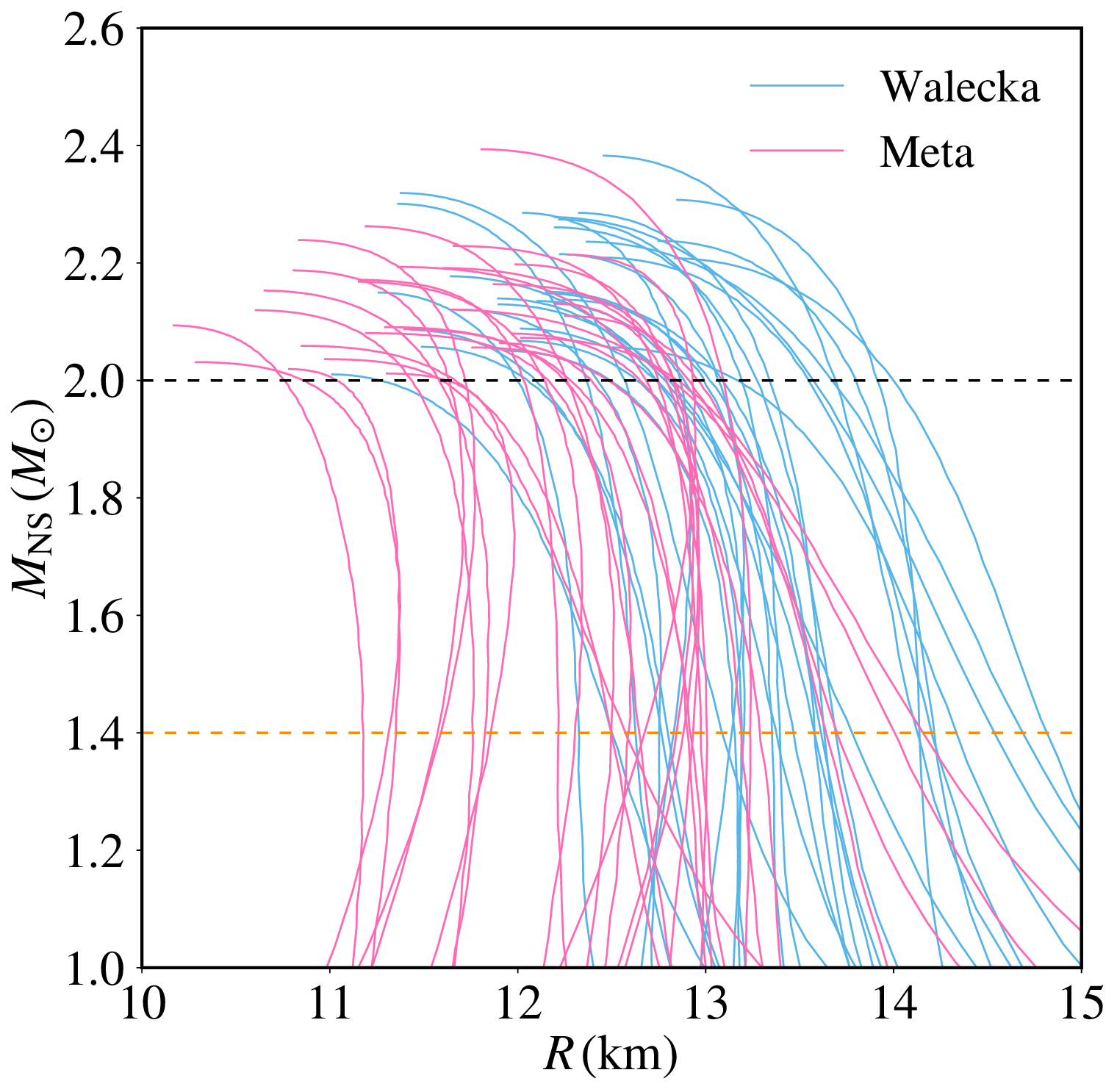}
\caption{(Color Online). NS M-R curves generated by the meta-model and Walecka model EOSs (60 sample are shown in the plot). The yellow (black) dashed line marks $M_{\mathrm{NS}}/M_\odot = 1.4$ ($M_{\mathrm{NS}}/M_\odot = 2.01$).
}\label{fig:MR}
\end{figure}

For the meta-model, we consider charge-neutral npe$\mu$ matter in $\beta$-equilibrium. The energy per nucleon $E(\rho,\delta)$ given by Eq.\,\eqref{eq:e-expansion} consists of the EOS of SNM and the symmetry energy $E_{\mathrm{sym}}(\rho)$, parameterized respectively by Eq.\,\eqref{eq:e0-expansion} and Eq.\,\eqref{eq:esym-expansion}. The $\beta$-equilibrium condition $\mu_\rm n-\mu_\rm p=\mu_\rm e=\mu_\mu \approx 4\delta E_{\mathrm{sym}}(\rho)$, together with the charge neutrality condition $\rho_\rm p=\rho_\rm e+\rho_\mu$, determines the isospin asymmetry $\delta=\delta(\rho)$ as a function of the baryon density, where the chemical potential of particle species $i$ is obtained from $\mu_i=\partial\varepsilon(\rho,\delta)/\partial\rho_i$. For the Walecka model, the full EOS of ANM can be obtained exactly, without invoking the parabolic approximation in evaluating $\mu_\rm n-\mu_\rm p$. Once the equilibrium composition is determined, the total energy density is calculated through $\varepsilon(\rho,\delta)=\rho[E(\rho,\delta)+M_{\rm N}]+\varepsilon_\ell$, where $\varepsilon_\ell$ denotes the total lepton energy density from electrons and muons treated as ideal Fermi gases\,\cite{Oppenheimer1939}. The pressure is then obtained from $P(\rho,\delta)=\rho^2\d[\varepsilon(\rho,\delta)/\rho]/\d\rho$. In this way, the two quantities $\varepsilon(\rho)$ and $P(\rho)$ are determined and the EOS $P(\varepsilon)$ is constructed.
The core-crust transition density $\rho_{\rm t}$ is determined using the thermodynamic method\,\cite{Xu2009,Kubis2007,Cai2012}. For the inner crust, with densities between $\rho_{\rm t}$ and $\rho_{\mathrm{out}}\approx2.46\times10^{-4}\,\mathrm{fm^{-3}}$, we adopt the parametrized EOS $P=\alpha+\beta\varepsilon^{4/3}$\,\cite{Iida1997,Zhang2016}. For the outer crust below $\rho_{\mathrm{out}}$, we employ the Baym--Pethick--Sutherland (BPS) and Feynman--Metropolis--Teller (FMT) EOSs\,\cite{Baym1971}.

For the Walecka-model sampling, in order to ensure a sufficiently broad parameter space, we sample the saturation density in the range $0.14\,\mathrm{fm^{-3}}\lesssim\rho_0\lesssim0.18\,\mathrm{fm^{-3}}$, the nucleon Dirac mass in SNM within $0.4\lesssim M_0^\ast/M_\rm N\lesssim0.8$, the binding energy per nucleon within $-18\,\mathrm{MeV}\lesssim E_0(\rho_0)\lesssim-14\,\mathrm{MeV}$, the incompressibility within $200\,\mathrm{MeV}\lesssim K_0\lesssim280\,\mathrm{MeV}$, and the skewness parameter within $-1000\,\mathrm{MeV}\lesssim J_0\lesssim0\,\mathrm{MeV}$. For the symmetry energy sector, we adopt $26\,\mathrm{MeV}\lesssim E_{\mathrm{sym}}(\rho_0)\lesssim38\,\mathrm{MeV}$ and $20\,\mathrm{MeV}\lesssim L\lesssim100\,\mathrm{MeV}$.
For the meta-model, we adopt the same sampling ranges for $\rho_0$, $E_0(\rho_0)$, $K_0$, $J_0$, $E_{\mathrm{sym}}$ and $L$ as in the Walecka model. In addition, for quantities not treated as independent inputs in the Walecka framework, we sample $300\,\mathrm{MeV}\lesssim I_0\lesssim600\,\mathrm{MeV}$, $-250\,\mathrm{MeV}\lesssim K_{\mathrm{sym}}\lesssim0\,\mathrm{MeV}$, and $200\,\mathrm{MeV}\lesssim J_{\mathrm{sym}}\lesssim1000\,\mathrm{MeV}$, see Refs.\,\cite{ZhangLi2018,ZhangLi2019,XieLi2020,ZhangLi2021,CaiLi2025Trace} and references cited therein for more details on these quantities.

\begin{figure}[h!]
\centering
\includegraphics[width=7.cm]{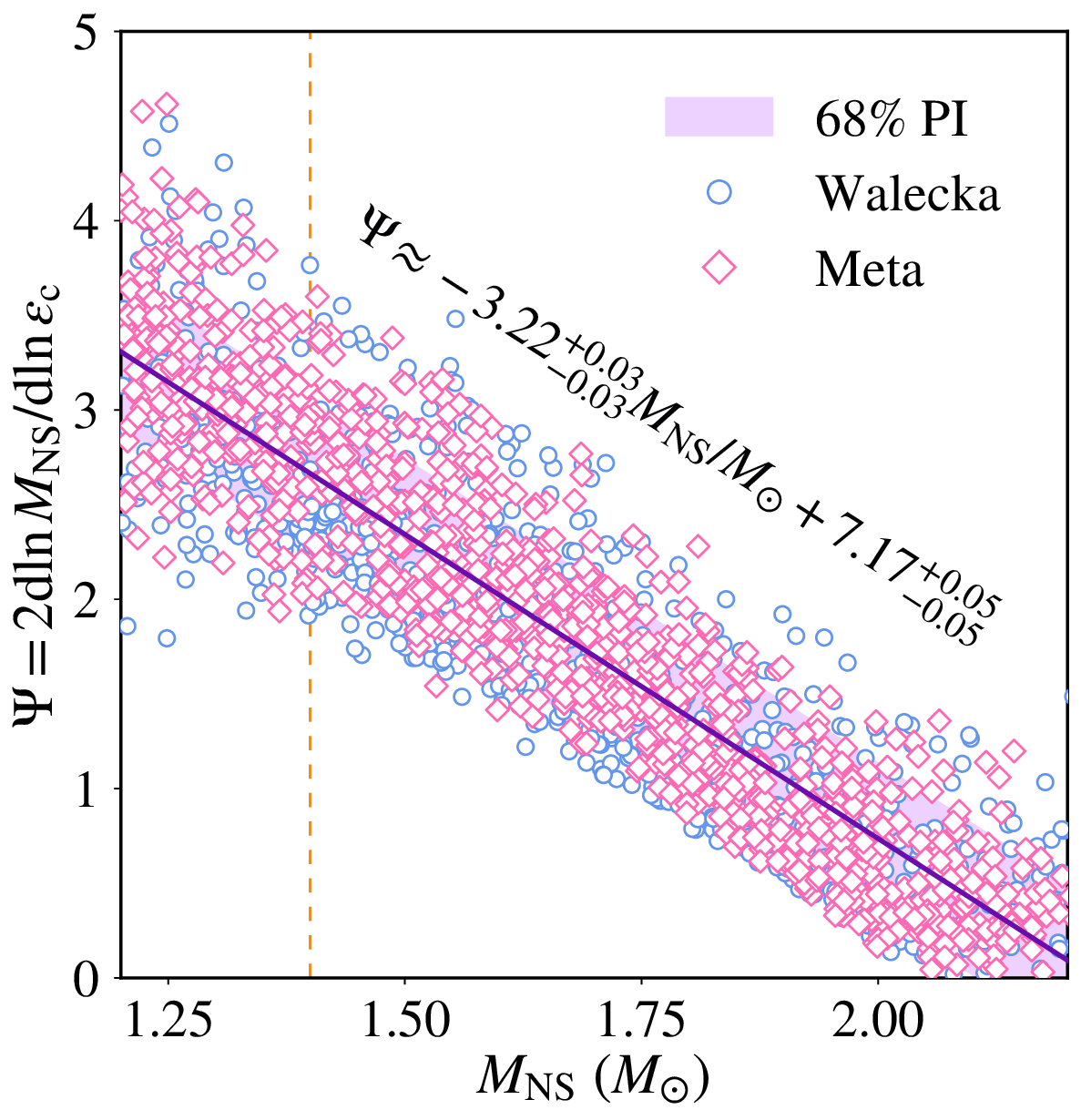}
\caption{(Color Online). The $\Psi$-$M_{\mathrm{NS}}$ correlation obtained from the $2\times10^3$ randomly sampled NS M-R curves generated by the meta-model EOSs and Walecka-model EOSs is shown. The yellow dotted line denotes $M_{\mathrm{NS}} = 1.4M_\odot$, corresponding to $\Psi \approx 2.66\pm0.40$.
}\label{fig:psi}
\end{figure}

For each sampled parameter set in the Walecka model, the coupling constants are inversely determined by solving the corresponding analytical equations, and these couplings are subsequently used to construct the EOS of NS matter. For the meta-model, each sampled parameter set is directly implemented in the expressions for $E_0(\rho)$ and $E_{\mathrm{sym}}(\rho)$ to construct the EOS accordingly. In addition, we require that the resulting EOS satisfies causality, and supports a maximum NS mass roughly larger than about $2.01M_\odot$, in order to be consistent with current astrophysical observations\,\cite{Demorest2010,
Antoniadis2013,Cromartie2020,Fon21,Miller21,
Riley21,Dittmann2024,Sullivan2024}. Finally, we collect $10^3$ valid parameter sets for each model and generate $10^3$ physically acceptable EOSs for NSs, the scaling properties converge quickly as the number of EOS samples reaches about several hundreds.

To illustrate the diversity of the generated EOSs, we randomly display 30 out of the $10^3$ M-R curves for each EOS class in FIG.\,\ref{fig:MR}. As shown in the figure, both the shapes of the curves and their slopes $\d M_{\mathrm{NS}}/\d R$ exhibit broad variations. For instance, the radius $R_{1.4}$ of a canonical NS spans approximately from 11\,km to 15\,km, indicating that the resulting M-R relations cover a wide and physically relevant EOS parameter space.

Another quantity closely related to our analysis is the log-stability slope $\Psi \equiv 2\d\ln M_{\mathrm{NS}}/\d\ln\varepsilon_{\rm c}$, which effectively characterizes the response of the NS mass to variations in the central energy density\,\cite{CaiLiMa2026phi}. This is because, for general stable NSs, the scaling factor $D(\x,\Psi)$ discussed in the main text depends not only on the central EOS-parameter $\x$, but also explicitly on $\Psi$, as shown in Eq.\,\eqref{eq:k2}. Using the $2\times10^3$ EOS samples generated from the Walecka model and the meta-model, we find the approximate relation
\begin{equation}
\Psi \approx -3.22^{+0.03}_{-0.03}\left(\frac{M_{\rm{NS}}}{M_\odot}\right)+7.17^{+0.05}_{-0.05}.
\label{eq:43}
\end{equation}
For instance, canonical NSs correspond to $\Psi\approx2.66\pm0.40$, implying a super-linear growth of the NS mass on the central energy density, namely $M_{\rm{NS}}\sim\varepsilon_{\rm c}^{1.33}$.

\begin{figure}[h!]
\centering
\includegraphics[width=8.cm]{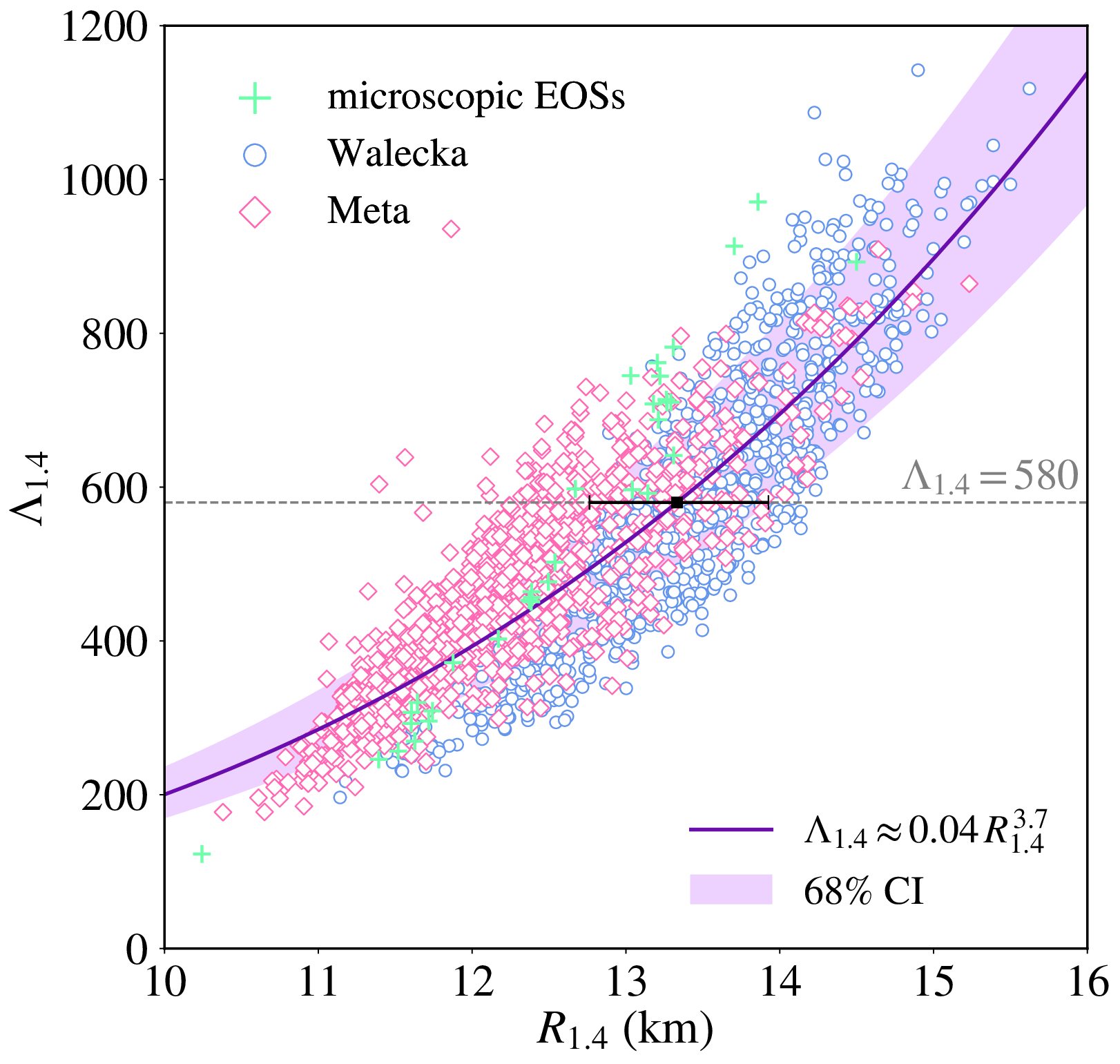}
\caption{(Color Online). The same as FIG.\,\ref{fig:Lambda14} but for $\Lambda_{1.4}$-$R_{1.4}$ correlation obtained from the $2\times10^3$ randomly sampled EOSs from the Walecka model and the meta model. The gray dashed line marks the upper bound $\Lambda_{1.4}\approx580$ from GW170817\,\cite{Abbott2018}.
}\label{fig:LambdaR}
\end{figure}

\section{Expressions for $H_1$ and $H_2$ of Eq.\,\eqref{eq:k2}}
\label{app:A}

The expressions for $H_1$ and $H_2$ appearing in Eq.\,\eqref{eq:k2} are defined as
\begin{align}
H_1=&2-Y+4\Pi_{\rm c}(Y-1),\\
H_2=&12\Pi_{\rm c}\left[
2-{Y}
+2\Pi_{\rm c}(5{Y}-8)
\right]
+
32\Pi_{\rm c}^3 {H},
\end{align}
where 
\begin{align}
    Y=&2-
\frac{2}{7}\frac{\x}{1+3\x}
\left(
\frac{1+36\x+99\x^2}{1+4\x+3\x^2}
+\frac{3}{s_{\rm c}^2}
\right),\\
H=&13-11Y
+2\Pi_{\rm c}(3{Y}-2)
+8\Pi_{\rm c}^2(1+{Y}),
\end{align}
and $s_{\rm{c}}^2$ and $\Pi_{\rm{c}}$ are defined in Eqs.\,\eqref{eq:sc2} and \eqref{eq:k2}, respectively.

\section{Correlation between $\Lambda_{1.4}$ and $R_{1.4}$}\label{app-c}

In FIG.\,\ref{fig:LambdaR}, we display the correlation between $\Lambda_{1.4}$ and $R_{1.4}$ for canonical NSs obtained from our $2\times10^3$ EOS samples. Such correlations have been widely used to infer the canonical radius $R_{1.4}$ from measured tidal deformabilities\,\cite{De18,Malik2018,ZhouChenZhang2019}. Using the effective parametrization $\Lambda_{1.4}\approx aR_{1.4}^b$, we extract the coefficients $a\approx0.04\pm 0.005 $ and $b\approx3.7\pm 0.05 $. Adopting the observational upper bound $\Lambda_{1.4}\lesssim580$ then yields $R_{1.4}\lesssim13.3\pm0.6\,\mathrm{km}$, consistent with previous analyses\,\cite{De18,Malik2018,ZhouChenZhang2019}.

We note, however, that the overall correlation shown in FIG.\,\ref{fig:LambdaR} exhibits a visibly larger scatter than the proposed $\Lambda$-$D$ scaling displayed in FIG.\,\ref{fig:Lambda14} and FIG.\,\ref{fig:LambdaTOV}. This indicates that the $\Lambda$-$D$ scaling established in the present work captures the underlying EOS dependence more tightly and therefore provides a more robust and efficient approach for constraining/extracting the EOS of dense matter in NSs.


\begin{references}

\bibitem{Abbott2017}B. Abbott et al. (LIGO Scientific Collaboration and Virgo Collaboration), \href{https://doi.org/10.1103/PhysRevLett.119.161101}{Phys. Rev. Lett. \textbf{119}, 161101 (2017).}

\bibitem{Abbott2018}B. Abbott et al. (LIGO Scientific Collaboration and Virgo Collaboration), \href{https://doi.org/10.1103/PhysRevLett.121.161101}{Phys. Rev. Lett. \textbf{121}, 161101 (2018).}


\bibitem{Riley19}T. Riley et al., \href{https://iopscience.iop.org/article/10.3847/2041-8213/ab481c}{Astrophys. J. Lett. \textbf{887}, L21 (2019).}

\bibitem{Miller19}M. Miller {et al.}, \href{https://iopscience.iop.org/article/10.3847/2041-8213/ab50c5}{Astrophys. J. Lett. \textbf{887}, L24 (2019).}


\bibitem{Fon21} E. Fonseca {et al.},  \href{https://iopscience.iop.org/article/10.3847/2041-8213/ac03b8}{Astrophys. J. Lett. \textbf{915}, L12 (2021).}

\bibitem{Riley21}T. Riley {et al.}, \href{https://iopscience.iop.org/article/10.3847/2041-8213/ac0a81}{Astrophys. J. Lett. \textbf{918}, L27 (2021).}

\bibitem{Miller21}M. Miller {et al.}, \href{https://iopscience.iop.org/article/10.3847/2041-8213/ac089b}{Astrophys. J. Lett. \textbf{918}, L28 (2021).}

\bibitem{Salmi22}T. Salmi et al., \href{https://iopscience.iop.org/article/10.3847/1538-4357/ac983d/pdf}{Astrophys. J. \textbf{941}, 150 (2022).}

\bibitem{Choud24}D. Choudhury et al., \href{https://iopscience.iop.org/article/10.3847/2041-8213/ad5a6f}{Astrophys. J.  Lett. \textbf{971}, L20 (2024).}

\bibitem{Reardon24}D. Reardon et al., \href{https://iopscience.iop.org/article/10.3847/2041-8213/ad614a}{Astrophys. J. Lett. \textbf{971}, L18 (2024).}

\bibitem{Mauv25}L. Mauviard et al., \href{https://iopscience.iop.org/article/10.3847/1538-4357/ae145d}{Astrophys. J. \textbf{995}, 60 (2025).}

\bibitem{Baym18}G. Baym {et al.}, \href{https://iopscience.iop.org/article/10.1088/1361-6633/aaae14}{Rep. Prog. Phys. \textbf{81}, 056902 (2018).}

\bibitem{Isa18}I. Vida$\widetilde{\rm{n}}$a, \href{https://royalsocietypublishing.org/doi/10.1098/rspa.2018.0145}{Proc. Roy. Soc. Lond. A \textbf{474}, 0145 (2018).}

\bibitem{LCXZ21}B.A. Li et al., \href{https://www.mdpi.com/2218-1997/7/6/182}{Universe \textbf{7}, 182 (2021).}

\bibitem{Latt21}J. Lattimer, \href{https://www.annualreviews.org/content/journals/10.1146/annurev-nucl-102419-124827}{Ann. Rev. Nucl. Part. Sci. \textbf{71}, 433 (2021).}

\bibitem{Oertel17} M. Oertel {et al.}, \href{https://journals.aps.org/rmp/abstract/10.1103/RevModPhys.89.015007}{Rev. Mod. Phys.  \textbf{89}, 015007 (2017).}

\bibitem{Bai19}L. Baiotti, \href{https://www.sciencedirect.com/science/article/pii/S0146641019300493}{Prog. Part. Nucl. Phys. \textbf{109}, 103714 (2019).}

\bibitem{Dri21}C. Drischler, J. Holt, and C. Wellenhofer, \href{https://www.annualreviews.org/doi/10.1146/annurev-nucl-102419-041903}{Annu. Rev. Nucl. Part. Sci. \textbf{71}, 403 (2021).}

\bibitem{Lovato22}A. Lovato {et al.},  \href{https://arxiv.org/abs/2211.02224}{arXiv:2211.02224 (2022).}

\bibitem{Soren23}A. Sorensen {et al.}, \href{https://doi.org/10.1016/j.ppnp.2023.104080}{Prog. Part. Nucl. Phys.  \textbf{134}, 104080 (2024).}

\bibitem{Chat24}K.  Chatziioannou et al., \href{https://journals.aps.org/rmp/abstract/10.1103/ymsq-cfcw}{Rev. Mod. Phys. \textbf{97}, 045007 (2025).}

\bibitem{Alar25}J. Alarcon, E. Lope-Oter, and Y. Cano, \href{https://arxiv.org/abs/2511.04737}{arXiv:2511.04737 (2025).}

\bibitem{LiA25}A. Li et al., \href{https://link.springer.com/article/10.1007/s11433-025-2761-4}{SCIENCE CHINA Physics, Mechanics and Astronomy \textbf{68}, 119503 (2025).}

\bibitem{Wat16}A. Watts et al., \href{https://journals.aps.org/rmp/abstract/10.1103/RevModPhys.88.021001}{Rev. Mod. Phys. \textbf{88}, 021001 (2016).}

\bibitem{LP01}J. Lattimer and M. Prakash, \href{https://iopscience.iop.org/article/10.1086/319702}{Astrophys. J. \textbf{550}, 426 (2001).}

\bibitem{Farr13}
F. Fattoyev et al., \href{https://journals.aps.org/prc/abstract/10.1103/PhysRevC.87.015806}{Phys. Rev. C \textbf{87}, no.1, 015806 (2013).}


\bibitem{De18}S. De et al., \href{https://journals.aps.org/prl/abstract/10.1103/PhysRevLett.121.091102}{Phys. Rev. Lett. \textbf{121}, 091102 (2018).}

\bibitem{Lim18}Y. Lim and J. Holt, \href{https://journals.aps.org/prl/abstract/10.1103/PhysRevLett.121.062701}{Phys. Rev. Lett. \textbf{121}, 062701 (2018).}

\bibitem{Zhao18}T. Zhao and J. Lattimer, \href{https://journals.aps.org/prd/abstract/10.1103/PhysRevD.98.063020}{Phys. Rev. D \textbf{98}, 063020 (2018).}

\bibitem{Most18}E. Most et al., \href{https://journals.aps.org/prl/abstract/10.1103/PhysRevLett.120.261103}{Phys. Rev. Lett. \textbf{120}, 261103 (2018).}

\bibitem{Zhou19}Y. Zhou and L.W. Chen, \href{https://iopscience.iop.org/article/10.3847/1538-4357/ab4adf/meta}{Astrophys. J. \textbf{886}, 52 (2019).}

\bibitem{Li2019}
B.A. Li et al.,
\href{https://link.springer.com/article/10.1140/epja/i2019-12780-8}{Eur. Phys. J. A \textbf{55}, no.7, 117 (2019).}

\bibitem{Chat20}K. Chatziioannou, \href{https://link.springer.com/article/10.1007/s10714-020-02754-3}{Gen. Relativ. Gravit. \textbf{52}, 109 (2020).}

\bibitem{Ferr24} M. Ferreira and C. Provid\^encia,
\href{https://journals.aps.org/prd/abstract/10.1103/PhysRevD.110.063018}
{Phys. Rev. D \textbf{110},  063018 (2024).}

\bibitem{Ferr25} M. Ferreira and C. Provid\^encia,
\href{https://journals.aps.org/prd/abstract/10.1103/r7gk-kcmn}
{Phys. Rev. D \textbf{112},  083058 (2025).}


\bibitem{ZhouX24}J. Zhou and J. Xu, \href{https://link.springer.com/article/10.1007/s11433-024-2406-4}{SCIENCE CHINA Physics, Mechanics and Astronomy \textbf{67}, 282011 (2024).}

\bibitem{Tsang24}C. Tsang et al., \href{https://www.nature.com/articles/s41550-023-02161-z}{Nature Astron. \textbf{8}, 328 (2024).}

\bibitem{Mroczek24}D. Mroczek et al., \href{https://journals.aps.org/prd/abstract/10.1103/PhysRevD.110.123009}{Phys. Rev. D \textbf{110}, 123009 (2024).}

\bibitem{Ripley24}J. Ripley et al., \href{https://www.nature.com/articles/s41550-024-02323-7}{Nature Astron. \textbf{8}, 1277 (2024).}

\bibitem{Malik24}T. Malik et al., \href{https://www.aanda.org/articles/aa/full_html/2024/09/aa49292-24/aa49292-24.html}{Astron. Astrophys. \textbf{689}, A242 (2024).}

\bibitem{CuiY25}Y. Cui et al. \href{https://link.springer.com/article/10.1007/s41365-025-01734-z}{Nucl. Sci. Tech. \textbf{36}, 141 (2025).}


\bibitem{Hot2011}
K. Hotokezaka et al.,
\href{https://journals.aps.org/prd/abstract/10.1103/PhysRevD.83.124008}{Phys. Rev. D \textbf{83}, 124008 (2011).}

\bibitem{Bau2012}
A.~Bauswein and H.~T.~Janka,
%``Measuring neutron-star properties via gravitational waves from binary mergers,''
\href{https://journals.aps.org/prl/abstract/10.1103/PhysRevLett.108.011101}{Phys. Rev. Lett. \textbf{108}, 011101 (2012).}

\bibitem{Rez2016}
L. Rezzolla and K. Takami,
%``Gravitational-wave signal from binary neutron stars: a systematic analysis of the spectral properties,''
\href{https://journals.aps.org/prd/abstract/10.1103/PhysRevD.93.124051}{Phys. Rev. D \textbf{93}, 124051 (2016).}

\bibitem{Bau2016}
A. Bauswein, N. Stergioulas and H. Janka, \href{https://link.springer.com/article/10.1140/epja/i2016-16056-7}{Eur. Phys. J. A \textbf{52}, 56 (2016).}

\bibitem{Mos2019}
E. Most et al., \href{https://journals.aps.org/prl/abstract/10.1103/PhysRevLett.122.061101}{Phys. Rev. Lett. \textbf{122}, 061101 (2019).}

\bibitem{CaiLi2025Review}
B.J. Cai and B.A. Li,
\href{https://doi.org/10.1140/epja/s10050-025-01507-7}{Eur. Phys. J. A \textbf{61}, 55 (2025).}

\bibitem{CaiLiZhang2023ApJ}
B.J. Cai, B.A. Li, and Z. Zhang,
\href{https://doi.org/10.3847/1538-4357/acdef0}{Astrophys. J. \textbf{952}, 147 (2023).}

\bibitem{CaiLiZhang2023PRD}
B.J. Cai, B.A. Li, and Z. Zhang,
\href{https://doi.org/10.1103/PhysRevD.108.103041}{Phys. Rev. D \textbf{108}, 103041 (2023).}

\bibitem{CaiLi2024SSS}
B.J. Cai and B.A. Li,
\href{https://doi.org/10.1103/PhysRevD.109.083015}{Phys. Rev. D \textbf{109}, 083015 (2024).}

\bibitem{CaiLi2025Trace}
B.J. Cai and B.A. Li,
\href{https://journals.aps.org/prd/accepted/87072Y45Gbc11185c8b90f097e5f10638ee20e84c}{Phys. Rev. D, \textbf{112}, 023023  (2025).}

\bibitem{Cai2024Front}
B.J. Cai and B.A. Li, 
\href{https://www.frontiersin.org/journals/astronomy-and-space-sciences/articles/10.3389/fspas.2024.1502888/full}{Front. Astron. Space Sci. \textbf{11}, 1502868 (2024).}

\bibitem{CaiLiMa2026phi}
B.J. Cai, B.A. Li, and Y.G. Ma,
\href{https://journals.aps.org/prd/abstract/10.1103/nbmw-k5fs}{Phys. Rev. D \textbf{113}, 023002 (2026).}

\bibitem{CaiLiMa2026X}
B.J. Cai, B.A. Li, and Y.G. Ma,
\href{https://arxiv.org/abs/2601.02980}{arXiv:2601.02980 (2026), Phys. Rev. D (2026) in press.}

\bibitem{Tolman1939}
R. Tolman,
\href{https://doi.org/10.1103/PhysRev.55.364}{Phys. Rev. \textbf{55}, 364 (1939).}

\bibitem{Oppenheimer1939}
J. Oppenheimer and G. Volkoff,
\href{https://doi.org/10.1103/PhysRev.55.374}{Phys. Rev. \textbf{55}, 374 (1939).}

\bibitem{Web99}F. Weber, \textit{Pulsars as Astrophysical Laboratories for Nuclear and Particle Physics}, Routledge, Chapter 14.


\bibitem{Hinderer2008}
T. Hinderer, 
\href{https://doi.org/10.1086/533487}{Astrophys. J. \textbf{677}, 1216 (2008).}

\bibitem{Fuji22}Y. Fujimoto {et al.}, \href{https://journals.aps.org/prl/abstract/10.1103/PhysRevLett.129.252702}{Phys. Rev. Lett. \textbf{129}, 252702 (2022).}

\bibitem{Saes24}J. Saes,  R. Mendes, and N. Yunes, \href{https://journals.aps.org/prd/abstract/10.1103/PhysRevD.110.024011}{Phys. Rev. D \textbf{110}, 024011 (2024).}

\bibitem{Marc24}M. Marczenko, \href{https://journals.aps.org/prc/abstract/10.1103/PhysRevC.110.045811}{Phys. Rev. C \textbf{110}, 04581  (2024).}

\bibitem{Li26}
B.A. Li, \href{https://journals.aps.org/prl/abstract/10.1103/b2t4-km2r}{Phys. Rev. Lett. \textbf{136}, 242301 (2026).}

\bibitem{SW1986}
B. Serot and J. Walecka,
\href{https://inspirehep.net/literature/207866}{Adv. Nucl. Phys. \textbf{16}, 1 (1986).}

\bibitem{ZhangLi2018}
N.B. Zhang, B.A. Li, and J. Xu,
\href{https://doi.org/10.3847/1538-4357/aac027}{Astrophys. J. \textbf{859}, 90 (2018).}

\bibitem{ZhangLi2019}
N.B. Zhang and B.A. Li,
\href{https://doi.org/10.3847/1538-4357/ab24cb}{Astrophys. J. \textbf{879}, 99 (2019).}

\bibitem{XieLi2020}
W.J. Xie and B.A. Li,
\href{https://doi.org/10.3847/1538-4357/aba271}{Astrophys. J. \textbf{899}, 4 (2020).}

\bibitem{ZhangLi2021}
N.B. Zhang and B.A. Li,
\href{https://doi.org/10.3847/1538-4357/ac1e8c}{Astrophys. J. \textbf{921}, 111 (2021).}

\bibitem{Typel15}S. Typel, M. Oertel, and T. Kl\"ahn, \href{https://doi.org/10.1134/S1063779615040061}{Phys. Part. Nucl. \textbf{46}, 633 (2015).}

\bibitem{Ofengeim24}D. Ofengeim, P. Shternin, and T. Piran, \href{https://doi.org/10.1103/PhysRevD.110.103046}{Phys. Rev. D \textbf{110}, 103046 (2024).}

\bibitem{Dong2025}
X. Dong et al.,
\href{https://doi.org/10.1103/hlgb-47pj}{Phys. Rev. D \textbf{112}, 063043 (2025).}

\bibitem{Golomb2025}
J. Golomb et al.,
\href{https://doi.org/10.1103/PhysRevD.111.023029}{Phys. Rev. D \textbf{111}, 023029 (2025).}

\bibitem{Saes21} J. Saes and R. Mendes, \href{https://arxiv.org/abs/2109.11571}{arXiv:2109.11571 (2021).}


\bibitem{JHChen24}J.H. Chen et al., \href{https://link.springer.com/article/10.1007/s41365-024-01591-2}{Nucl. Sci. Tech. \textbf{35}, 214 (2024).}


\bibitem{Ann23}E. Annala et al., \href{https://www.nature.com/articles/s41467-023-44051-y}{Nat. Comm. \textbf{14}, 8451 (2023).}

\bibitem{Sun2025}
B. Sun and J. Lattimer,
\href{https://doi.org/10.3847/1538-4357/adc25d}{Astrophys. J. \textbf{984}, 30 (2025).}

\bibitem{YagiYunes2013a}
K. Yagi and N. Yunes,
\href{https://www.science.org/doi/10.1126/science.1236462}{Science \textbf{341}, 365 (2013)}.

\bibitem{YagiYunes2013b}
K. Yagi and N. Yunes,
\href{https://journals.aps.org/prd/abstract/10.1103/PhysRevD.88.023009}{Phys. Rev. D \textbf{88}, 023009 (2013)}.

\bibitem{LCK08} B.A. Li, L.W. Chen, and C.M. Ko, \href{https://www.sciencedirect.com/science/article/pii/S0370157308001269}{Phys. Rep. \textbf{464}, 113 (2008).}

\bibitem{AnR24}R. An et al., \href{https://link.springer.com/article/10.1007/s41365-024-01551-w}{Nucl. Sci. Tech. \textbf{35}, 182 (2024).}

\bibitem{Ding24}M.Q. Ding, D.Q. Fang, and Y.G. Ma, \href{https://link.springer.com/article/10.1007/s41365-024-01584-1}{Nucl. Sci. Tech., 35, \textbf{211} (2024).}

\bibitem{LiuYY25}Y.Y. Liu et al., \href{https://link.springer.com/article/10.1007/s41365-024-01607-x}{Nucl. Sci. Tech. \textbf{36}, 45 (2025).}

\bibitem{Wang25}J.M. Wang et al., \href{https://iopscience.iop.org/article/10.1088/1674-1137/adf4a1}{Chinese Phys. C \textbf{49}, 124105 (2025).}

\bibitem{Cai17J0}B.J. Cai and L.W. Chen, \href{https://link.springer.com/article/10.1007/s41365-017-0329-1}{Nucl. Sci. Tech. \textbf{28}, 185 (2017).}

\bibitem{Mueller1996}
H. Müller and B. D. Serot,
\href{https://doi.org/10.1016/0375-9474(96)00187-X}{Nucl. Phys.  \textbf{A606}, 508 (1996).}

\bibitem{Li2022}
F. Li et al.,
\href{https://doi.org/10.3847/1538-4357/ac5e2a}{Astrophys. J. \textbf{929}, 183 (2022).}

\bibitem{ToddRutel2005}
B. Todd-Rutel and J. Piekarewicz,
\href{https://doi.org/10.1103/PhysRevLett.95.122501}{Phys. Rev. Lett. \textbf{95}, 122501 (2005).}

\bibitem{WeiSN24}S.N. Wei and Z.Q. Feng, \href{https://link.springer.com/article/10.1007/s41365-024-01380-x}{Nucl. Sci. Tech. \textbf{35}, 15 (2024).}


\bibitem{Cai2014}
B.J. Cai and L.W. Chen,
\href{https://doi.org/10.1007/s41365-017-0329-1}{Nucl. Sci. Tech. \textbf{28}, 185 (2017).}

\bibitem{Qin25}P.P. Qin et al., \href{https://link.springer.com/article/10.1007/s41365-024-01609-9}{Nucl. Sci. Tech. \textbf{36}, 29 (2025).}



\bibitem{Xu2009}
J. Xu et al.,
\href{https://doi.org/10.1088/0004-637X/697/2/1549}{Astrophys. J. \textbf{697}, 1549 (2009).}

\bibitem{Kubis2007}
S. Kubis,
\href{https://doi.org/10.1103/PhysRevC.76.025801}{Phys. Rev. C \textbf{76}, 025801 (2007).}

\bibitem{Cai2012}
B.J. Cai and L.W. Chen,
\href{https://doi.org/10.1103/PhysRevC.85.024302}{Phys. Rev. C \textbf{85}, 024302 (2012).}

\bibitem{Iida1997}
K. Iida and K. Sato,
\href{https://doi.org/10.1086/303685}{Astrophys. J. \textbf{477}, 294 (1997).}

\bibitem{Zhang2016}
Z. Zhang and L.W. Chen,
\href{https://doi.org/10.1103/PhysRevC.94.064326}{Phys. Rev. C \textbf{94}, 064326 (2016).}

\bibitem{Baym1971}
G. Baym, C. Pethick, and P. Sutherland,
\href{https://doi.org/10.1086/151216}{Astrophys. J. \textbf{170}, 299 (1971).}

\bibitem{Demorest2010}
P. Demorest et al.,
\href{https://doi.org/10.1038/nature09466}{Nature \textbf{467}, 1081 (2010).}

\bibitem{Antoniadis2013}
J. Antoniadis et al.,
\href{https://doi.org/10.1126/science.1233232}{Science \textbf{340}, 1233232 (2013).}

\bibitem{Cromartie2020}
H. Cromartie et al.,
\href{https://doi.org/10.1038/s41550-019-0880-2}{Nat. Astron. \textbf{4}, 72 (2020).}

\bibitem{Dittmann2024}
A. Dittmann et al.,
\href{https://doi.org/10.3847/1538-4357/ad5f1e}{Astrophys. J. \textbf{974}, 295 (2024).}

\bibitem{Sullivan2024}
A. Sullivan and R. Romani,
\href{https://doi.org/10.3847/1538-4357/ad4d85}{Astrophys. J. \textbf{974}, 315 (2024).}

\bibitem{Malik2018}
T. Malik et al.,
\href{https://doi.org/10.1103/PhysRevC.98.035804}{Phys. Rev. C \textbf{98}, 035804 (2018).}

\bibitem{ZhouChenZhang2019}
Y. Zhou, L.W. Chen, and Z. Zhang,
\href{https://doi.org/10.1103/PhysRevD.99.121301}{Phys. Rev. D \textbf{99}, 121301(R) (2019).}

\end{references}
\end{document}